\documentclass[acmtog]{acmart}

\usepackage{booktabs} % For formal tables
\usepackage{gensymb}
\usepackage{mystyle}
\usepackage{wrapfig}

\setcitestyle{square}

\newcommand{\revisedtext}[1]{{#1}}
\newcommand{\minorrev}[1]{{#1}}

\usepackage[ruled]{algorithm2e} % For algorithms

\SetAlFnt{\small}
\SetAlCapFnt{\small}
\SetAlCapNameFnt{\small}
\SetAlCapHSkip{0pt}
\IncMargin{-\parindent}

\acmJournal{TOG}
\acmVolume{0}
\acmNumber{0}
\acmArticle{0}
\acmYear{2017}
\acmMonth{0}

\setcopyright{acmlicensed}
\acmDOI{0000001.0000001_2}

\begin{document}
% Title portion
\title{Vectorization of Line Drawings via PolyVector Fields} 
\author{Mikhail Bessmeltsev}
\email{bmpix@mit.edu}
\affiliation{%
	\institution{Universit\'e de Montr\'eal \& Massachusetts Institute of Technology}
}
\author{Justin Solomon}
\email{jsolomon@mit.edu}
\affiliation{%
\institution{Massachusetts Institute of Technology}
}
% If accepted put authors here
\renewcommand\shortauthors{} % if accepted, put abbreviation here for authors

\begin{abstract}
% !TEX root = ../polyvectorization.tex

Image tracing is a foundational component of the workflow in graphic design, engineering, and computer animation, linking hand-drawn concept images to collections of smooth curves needed for geometry processing and editing. Even for clean line drawings, modern algorithms often fail to faithfully vectorize junctions, or points at which curves meet; this produces vector drawings with incorrect connectivity.  This subtle issue undermines the practical application of vectorization tools and accounts for hesitance among artists and engineers to use automatic vectorization software. To address this issue, we propose a novel image vectorization method based on state-of-the-art mathematical algorithms for frame field processing.  Our algorithm is tailored specifically to disambiguate junctions without sacrificing quality. 

\end{abstract}

%
% The code below should be generated by the tool at
% http://dl.acm.org/ccs.cfm
% Please copy and paste the code instead of the example below. 
%


%
% End generated code
%

\thanks{}

\begin{teaserfigure}
	\includegraphics[width=\textwidth]{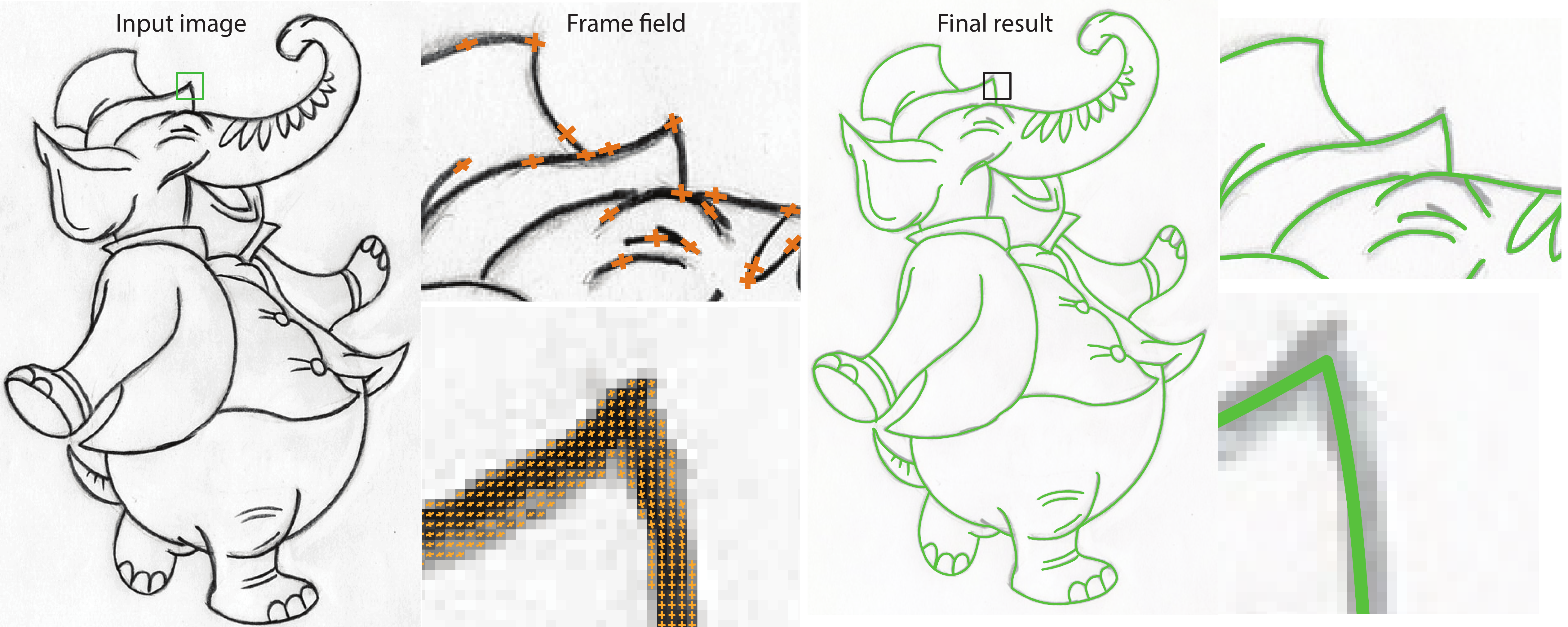}
	\caption{Given a possibly noisy grayscale bitmap image, we compute a frame field aligned with the directions on the image, superimposing multiple directions around sharp corners as well as X- and T-junctions. We then use this frame field to extract the drawing topology and create the final vectorization with the computed topology. Frame field computation (shown for a subset of pixels in the upper zoom and the full field in the lower one) is the key component of the system. The frame field disambiguates X- and T-junctions even in the noisy areas, allowing tracing to be straightforward and robust. Input images are from www.easy-drawings-and-sketches.com, \textcopyright Ivan Huska.}
	\label{fig:teaser}
\end{teaserfigure}

\maketitle

% !TEX root = ../polyvectorization.tex

\section{Introduction}

Image vectorization algorithms date back to early 1990s and are among the core tools in vector processing software including Adobe Illustrator (Live Trace), CorelDRAW (PowerTRACE), and Inkscape. Despite their wide adoption in industry, algorithms for line drawing vectorization remain under active development and still admit major shortfalls \cite{Noris2013,Favreau2016}. In several industries where vectorization is heavily needed, \revisedtext{including traditional animation and engineering design, this task frequently is} done manually, by painstakingly tracing a scanned image with drawing tools. This process is often considered to take less time than editing the automatic result from commercial vectorization tools. 

A primary reason for frustration with line drawing vectorization algorithms is incorrect treatment of junctions, resulting in wrong topology, or connectivity (Fig.\ref{fig:favreau_noris}(a,b)). Image understanding and perception rely on junctions and drawing topology to disambiguate depth and other cues~\cite{Xia2014}. In industries such as character animation, incorrect topology yields temporal incoherence and makes modern automatic coloring or in-betweening tools unusable~\cite{Whited2010,Orzan2013}. In engineering-oriented industries, incorrect topology may be considered an incorrect result overall, because it may not correspond to a physically-realizable object.
 
The main challenge when disambiguating junctions in line drawings is noisy or insufficient local information, even for clean images~\cite{Noris2013}. The presence of noise, such as uneven curve edges, complicates matters even further, and the widely-used local approach to resolving junctions based on a one-pixel width image skeleton \revisedtext{becomes} unreliable~\cite{Favreau2016}. 

A recent method by Favreau et al.~\shortcite{Favreau2016} uses \emph{global information} to resolve ambiguities at junctions. Their method successfully vectorizes sketches with numerous overdrawn strokes, where a heavy simplification of the result is needed. Unfortunately, for inputs requiring fidelity, their approach can lead to oversimplified results significantly deviating from the drawn contours (Fig.~\ref{fig:favreau_noris}(b),~\ref{fig:comparison_favreau}). 
  
While theoretically junctions may have various valences, as noted by previous work~\cite{Noris2013}, the vast majority of junctions
are X- and T-junctions (Fig.~\ref{fig:teaser}). Occlusion contours typically generate T-junctions, making them crucial for 3D shape perception~\cite{kanizsa1979organization,Bessmeltsev2015}. Hence, correct resolution of X- and T-junctions is a primary concern during image vectorization.

With these challenges in mind, in this paper we propose a robust image tracing method true to the image in unambiguous regions,  with global treatment of T- and X-junctions even when local information is unclear (Fig.~\ref{fig:favreau_noris}, right). 
Our technical innovation is to use frame fields to guide vectorization.  Frame fields attach two pairs of vectors $\{\pm u,\pm v\}$ to each point on the plane. They have been recently used to generate anisotropic quadrilateral meshes and to estimate 3D normals from 2D sketches~\cite{Panozzo2014,Iarussi2015}. Although frame fields are natural for tracking the orientations of curves meeting at sharp junctions, to our knowledge they never have been applied to image vectorization. 

\paragraph*{Overview.} As illustrated in Figure~\ref{fig:teaser}, the general idea of our method is to find a smooth frame field on the image plane, where at least one direction is aligned with nearby contours of the drawing. Around X- or T-shaped junctions, the two directions of the field will be aligned with the two intersecting contours. Then, we extract the topology of the drawing by tracing the frame field and grouping traced curves into strokes. Finally, we create a vectorization aligned with the frame field with the extracted topology.

% !TEX root = ../polyvectorization.tex

\section{Related Work}

\begin{figure}
	\includegraphics[width=0.9\columnwidth]{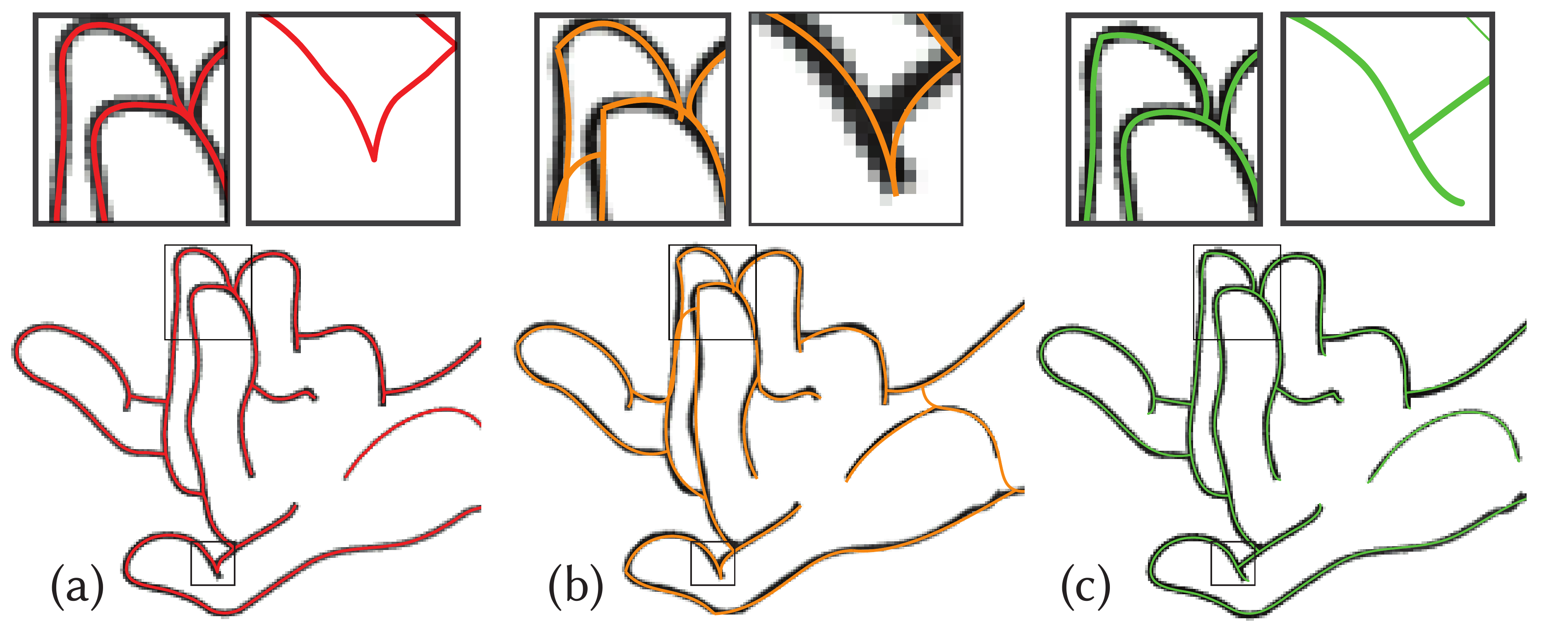}
	\caption{(a) Local approaches to junction resolution, such as the one proposed by Noris et al.~\shortcite{Noris2013}, may result in incorrect or imprecise junctions. (b) Favreau et al.'s method~\shortcite{Favreau2016} may significantly deviate from the drawing. (c) Our result.}
	\label{fig:favreau_noris}
\end{figure}

\begin{figure}
	\includegraphics[width=0.9\columnwidth]{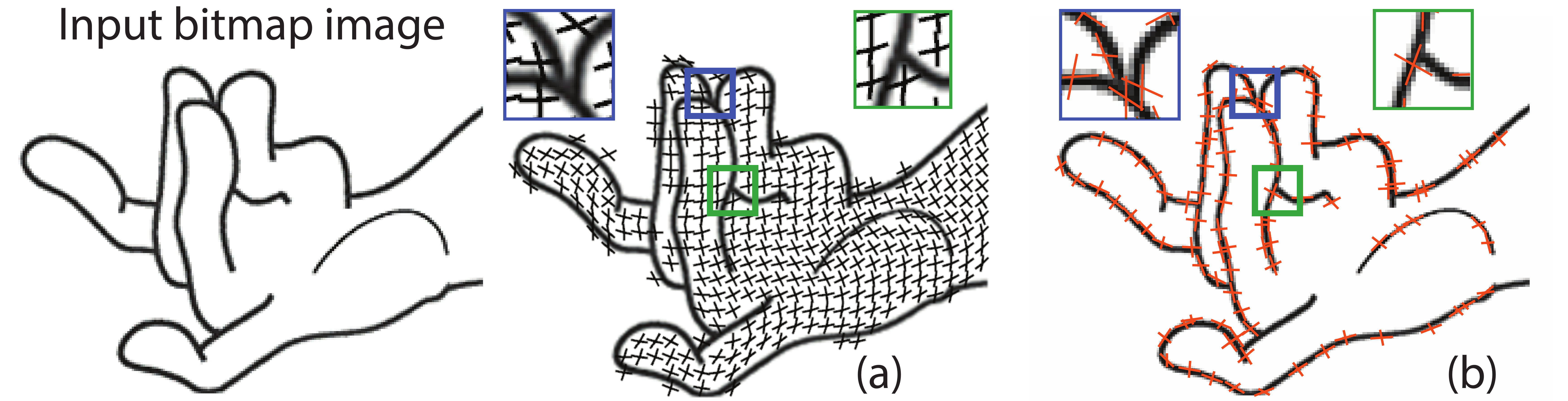}
	\caption{The target application for Bendfields \cite{Iarussi2015} leads to different frame field design assumptions (a) that are unsuitable for our type of vectorization. Compare with our result (b). For clarity, we only show a subset of frame field on (b); we compute frame field on every dark pixel.}
	\label{fig:bendfields}
\end{figure}

Our work builds upon achievements in three areas: image vectorization, junction detection, % in computer vision,  %<--- also in other areas
and frame fields. A comprehensive review of these areas is outside the scope of this paper; here, we instead highlight work relevant to our proposed pipeline.%focus on only the most relevant works. 

\paragraph*{Frame fields.} Our algorithm is built upon the construction of \emph{frame fields} that assign two directions to every point in a planar region; these directions will guide our placement of strokes.  Unlike \emph{cross fields}, frame fields have no constraint on orthogonality or length of the direction vectors.  We refer the reader to the recent survey by Vaxman et al.~\shortcite{Vaxman2016} for broad discussion.

While cross fields have been extensively used in computer graphics~\cite{Kass1987,Hertzmann2000,Palacios2007}, representations of frame fields and algorithms for their computation are relatively recent~\cite{Panozzo2014,Diamanti2015}. %Frame fields are a non-orthogonal and not-unit length generalization of cross-fields. % <--- Justin moved to short paragraph on top citing survey
They serve as a natural representation of linear transformations on tangent spaces of a surface. Frame fields originally were proposed for guiding anisotropic quad meshing via inversion-free mesh parameterization~\cite{Panozzo2014}.  Since then, frame fields have found additional applications, such as inferring 3D normals from a 2D sketch~\cite{Iarussi2015} and recovery of damaged historical documents \cite{Pal16Digitally}. 

Our work is driven by the frame field synthesis and interpolation tool set developed in~\cite{Panozzo2014,Diamanti2015}. Namely, we use their definition and representation of a \textit{PolyVector field}, as described in Section \ref{sec:frame_fields}.

%In frame field applications, we  are 
The BendFields algorithm proposed by Iarussi et al.~\shortcite{Iarussi2015} inspired some aspects of our approach.  While their algorithm is targeted to 3D surface reconstruction from curvature lines, they initially generate a frame field aligned to directions in a bitmap image. Their goal, however, is to compute a frame field in the space \emph{between} the input curves, while we solve for a frame field defined exclusively on dark pixels.  This difference gives our method a significant performance boost by reducing the number of degrees of freedom, and it qualitatively affects the results near junctions with sharp angles. Due to differences in application, the formulation and weighting of their alignment term differs from ours (Fig.~\ref{fig:bendfields}), and our use of PolyVectors has only real-valued variables per pixel rather than requiring a mixed-integer solver.

\paragraph*{Image vectorization.} Vectorization of bitmapped images has been studied extensively in graphics, vision, and other disciplines. Various input- and application-specific priors guide many vectorization methods, conforming to requirements of end users in medical imaging, road map reconstruction from GPS traces, processing of astronomical imagery, and other tasks~\cite{Turetken2013,Chai2013,Bo2016}. These methods are application-specific and cannot be applied directly to vectorization of hand-drawn line drawings.  Other vectorization methods deal with shaded images, like photographs or cartoon images~\cite{Zhang2009,Lecot2006,Orzan2013}; their focus is to capture an image using simple colored primitives, which typically are assumed to be closed. 

We focus on reconstruction of line drawings without shading, where lines may or may not be closed. In this area, strong priors about line shape, e.g.\ that lines only form circles or straight lines, %such as limiting possible shapes to circles and straight lines, can 
might bring simplicity to vectorization of technical drawings~\cite{Hilaire2006}, but do not apply to free-form line drawings. 

For vectorization of curvy line drawings, existing methods vary by the amount of noise allowed in the input. Noisy line drawings with multiple overlapping strokes or hatching patterns require deviation %vectorization methods to deviate 
from the drawn image in favor of simplicity~
% of the drawing 
\cite{Bartolo2007,Favreau2016}. Guided by a similar motivation, De Goes et al.~\shortcite{DeGoes2014} propose a method to extract a simplified curve network for noisy drawings. While their approach is natural for drawings with very fuzzy lines and significant noise, such behavior may not be desired for higher-quality drawings that do not contain overlapping strokes, which require more precise vectorization (Fig.\ \ref{fig:favreau_noris}(a)). 

On the other side of the spectrum is an image vectorization method tailored for clean cartoon drawings by Noris et al.~\shortcite{Noris2013}. Their global approach to topology allows them to, for instance, correctly disambiguate nearby parallel strokes.
Their treatment of junctions, however, is still local and may result in incorrect or imprecise treatment (Fig. \ref{fig:favreau_noris}(b)). Furthermore, the discrete nature of the algorithm renders it unstable in presence of noise (Fig. \ref{fig:comparison_noris}).

\begin{figure}
	\includegraphics[width=0.9\columnwidth]{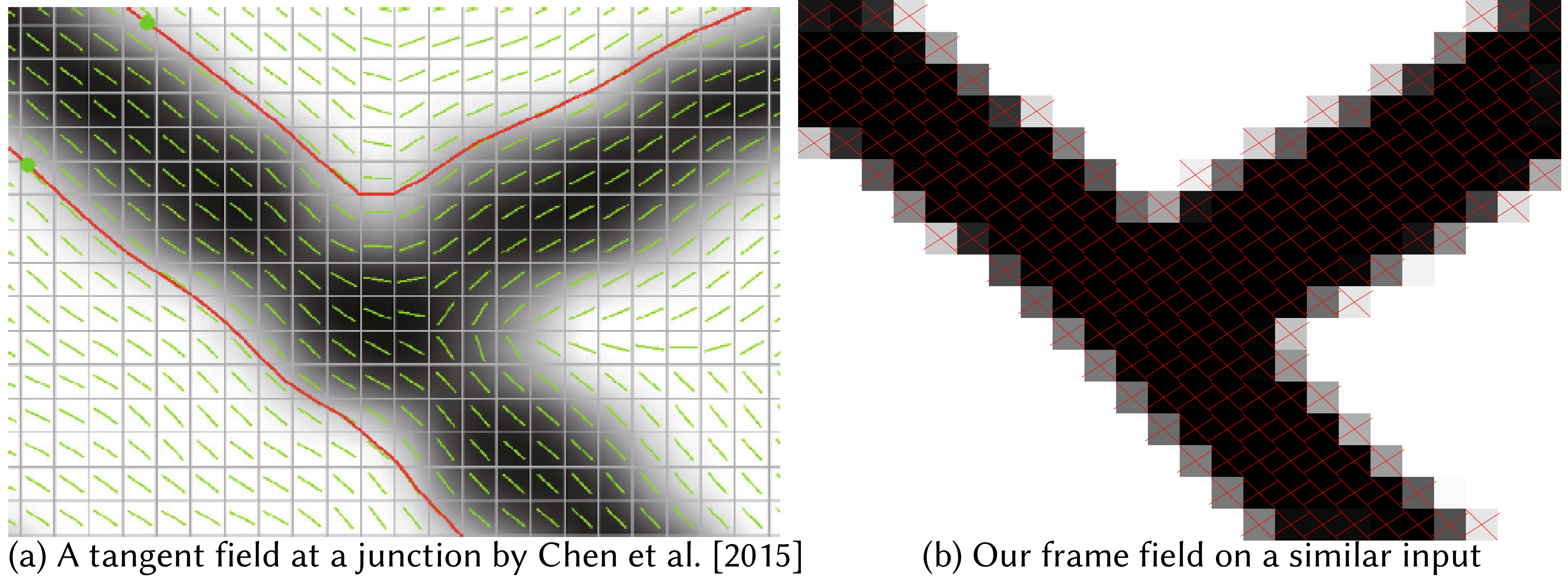}
	\caption{(a) A tangent field \cite{Chen2015} cannot capture a collection of directions present at a junction point. (b) On a similar input, the frame field is a natural representation of the directions at a junction.}
	\label{fig:comparison_chen}
\end{figure}

A recent work by Donati et al. \shortcite{Donati_2017_ICCV} explores accurate vectorization of noisy sketches using Pearson's correlation coefficient with Gaussian kernels. While their method achieves impressive performance and is able to process sketches with multiple overlapping strokes, it makes no effort to correctly disambiguate junctions, parallel lines, or overall extract drawing topology. Instead they rely on the topology of a 1-pixel width skeleton, which is known to be prone to local artifacts \cite{Favreau2016}. In contrast, we resolve junctions and parallel lines by generating a frame field, and use it to explicitly extract drawing topology. 

A line of work close to our method is using tangent fields for image processing and vectorization \cite{chen:hal-00857265, Kang2007, Chen2015}. For instance, Chen et al.~\shortcite{Chen2015} propose an image vectorization method with a global variational approach to disambiguation of junctions. The primary issue with these approaches is the use of tangent fields, which cannot capture a \emph{collection} of directions present at a junction point (Fig \ref{fig:comparison_chen}). As a result, the method by Chen et al. \shortcite{Chen2015} relies on user interaction and arbitrary thresholds to resolve junctions. Their method also does not consider the topology of the drawing, potentially yielding disconnected lines and/or spurious connections. %\misha{I would maybe remove the last sentence: we are making enough of a point maybe?} \justin{How about:  Moreover, their tracing technique does not consider global topology.}

Building on this work, our method uses a more natural representation to track junctions in the drawings:  a frame field defined at each stroke pixel. The two directions of the frame field %allow us to 
efficiently disambiguate directions around T- or X-junctions, and the variational nature of our approach makes it resistant to %normal % <--- like "surface normal" ?
noise. 

\paragraph*{Corner and junction detection in images.}  Corner detection is a basic step in classical computer vision pipelines \cite{Szeliski2010}; for example, the well-known Harris corner detector \cite{Harris88acombined} is implemented in countless industry-standard vision libraries.  The goal of these methods typically is to detect and characterize salient features, while for vectorization it is more important to calculate the exact center of the junction and to estimate the directions of the joining lines robustly. Furthermore, even if it is possible to identify junction points, image gradient directions near corners and junctions often are noisy, making it difficult to estimate the individual directions meeting at a junction point using purely local information.

% !TEX root = ../polyvectorization.tex

\section{Algorithm}

 \begin{figure}
	\includegraphics[width=\linewidth]{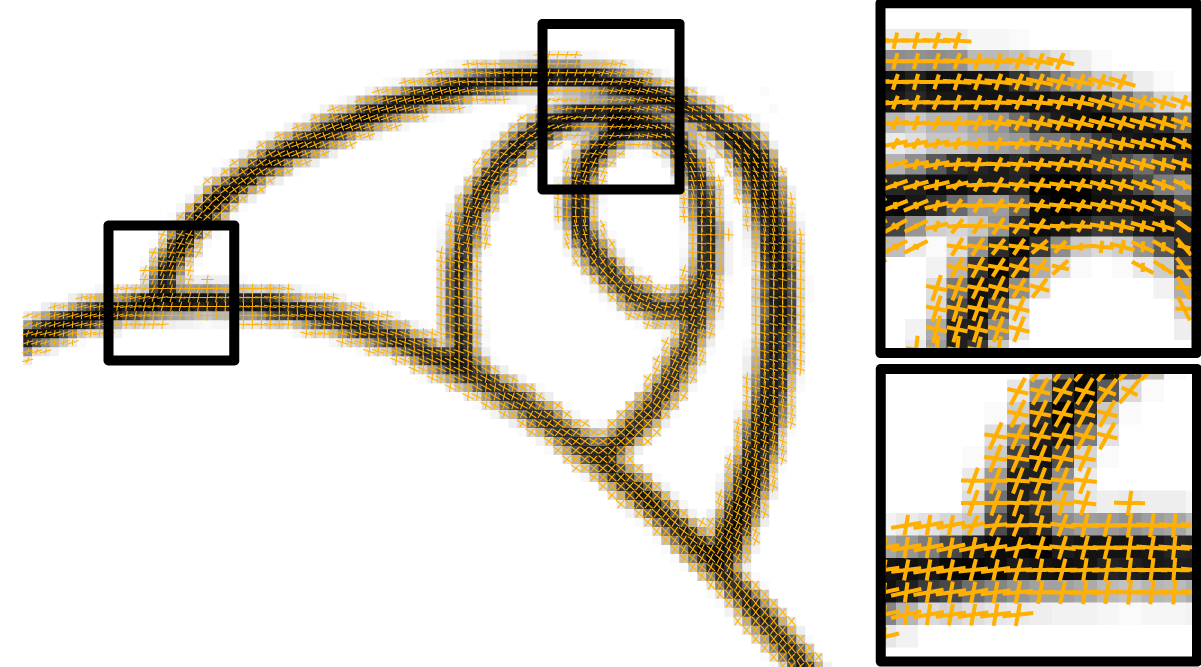}
	\caption{The frame field we design has at least one field direction aligned with a nearby curve tangent. Near T- and X-junctions, our field is aligned with both tangent directions.}
	\label{fig:frame_field}
\end{figure}

Our system takes as input a grayscale bitmap line drawing and produces a set of strokes aligned to the drawing.  We first solve an optimization problem to compute a frame field at each pixel in a narrow band around the set of stroke pixels, designed to capture directionality of the input and to superpose multiple directions near junctures.  We then extract topology of the drawing by tracing the frame field and grouping curves into strokes. We then compute the final vectorization (Fig. \ref{fig:teaser}). 

\subsection{Designing Frame Fields}
\label{sec:frame_fields}

Our vectorization algorithm begins by computing a smooth frame field, such that at every point near a stroke at least one field direction is aligned with a nearby curve tangent.  Near T- and X-junctions, our field will align to \emph{both} tangent directions present nearby in the image; this provides the flexibility needed to resolve image behavior near junctions (Fig. \ref{fig:frame_field}). 
%Naturally, such a field would align with both directions around T- or X-junctions, which will allow us to effectively disambiguate the junction regions. 
We formulate computation of the field as a variational problem that, after discretization, can be approached using standard algorithms for nonlinear unconstrained optimization. 

\paragraph*{Initial steps.}  We start by thresholding the image to isolate those pixels involved in the line drawing.  In particular, our algorithm will operate on a subset of pixels $I$, corresponding to dark pixels with intensity less than a fixed threshold, \revisedtext{$\theta_{\textrm{noise}} \cdot I_{max}$,} where $I_{max}$ is the maximum image intensity. We additionally estimate a noisy tangent field $\tau$ from the drawing by taking its Sobel gradient $g$ \revisedtext{with kernel of size 3} and rotating by $\nicefrac{\pi}{2}$.  Note that, \revisedtext{similarly to the discussion in \cite{Zhang2007}}, the direction of this rotation (clockwise vs.\ counterclockwise) will not affect the computation of the frame field, which always couples forward and backward directions. 
 
\paragraph*{Representation and variational problem.}
Following Diamanti et al.~\shortcite{Diamanti2015}, we represent the unknown frame field as a \emph{PolyVector field}.  Suppose we are given two directions $u,v$ representing curve tangents of the drawing near a given pixel; we can identify the image plane with the complex numbers $\C$ and take $u,v\in \C$ as complex vectors.  Consider the following complex polynomial $f(z)$:
%at each point as roots of complex polynomials
  \begin{equation}
  \label{eq:complex_poly}
  f(z)\eqdef (z^2-u^2)(z^2-v^2)=z^4  + c_2  z^2  + c_0.
  \end{equation}
Here, the constants $c_0$ and $c_2$ determine $u$ and $v$ up to relabeling and sign.  That is, every pair $(c_0,c_2)\in\C^2$ uniquely determines a \emph{frame} $\{\pm u,\pm v\}$, agnostic to the labeling of $u$ vs.\ $v$ as well as their sign.  We use $f(z;c_0,c_2)$ to denote the function above parameterized by the two coefficients $c_0$ and $c_2$.

Recovering the frame directions from $(c_0,c_2)$ is equally straightforward:
\begin{equation}\label{eq:coefftoframe}
\left\{
\begin{array}{r@{\,}l}c_0&=u^2v^2\\c_2&=-(u^2+v^2)\end{array}
\right\}
\longleftrightarrow
\left\{
\begin{array}{r@{\,}l}
u^2 &=-\frac{1}{2}\left(c_2+\sqrt{c_2^2-4c_0}\right)\\
v^2 &=-\frac{1}{2}\left(c_2-\sqrt{c_2^2-4c_0}\right)
\end{array}
\right\}
\end{equation}
Of course, the relationship on the right is non-unique.
  
Optimizing for a $(u,v)$ pair per pixel induces challenging issues involving labeling and sign; for example, this representation in the BendFields algorithm requires the use of a mixed-integer solver~\cite{Iarussi2015}.  To avoid this complexity, we instead optimize for a $(c_0,c_2)$ pair per pixel, which has no sign or ordering ambiguity. That is, the unknown in our optimization technique is a pair of complex-valued functions $c_0,c_2: I\rightarrow\C$.

We propose the following variational problem:% for recovering these functions:
%\begin{equation}
%\label{eq:variational}
%\begin{split}
\begin{align}
\min_{c_0,c_2:I \rightarrow \C} & E_{\mathrm{alignment}} + \lambda E_{\mathrm{smoothness}} + \mu E_{\mathrm{regularization}} = \nonumber\\
\min_{c_0,c_2:I \rightarrow \C} & \int_I{|f(e^{i\theta_\tau};c_0 (\vec{x}),c_2 (\vec{x}))|^2\,d\vec{x}+\lambda \sum_{i=0,2} \int_I\|\nabla c_i (\vec{x})\|^2\,d\vec{x}} \nonumber\\
	& + \mu \int_I|f(e^{i\theta_g};c_0 (\vec{x}),c_2 (\vec{x}))|^2\,d\vec{x}\label{eq:variational}
\end{align}
%\end{split}
%\end{equation}
Here, $\theta_w$ encodes the direction of a vector $w$, i.e.\ in complex language we can write $w=\|w\|_2e^{i\theta_w}$.

Details about the individual terms in~\eqref{eq:variational} are below, in the order they appear:
\begin{itemize}[leftmargin=*]
\item \textsc{Alignment:}  The first optimization term enforces alignment of the frame field with the tangent directions.  This term is small when the polynomial $f(\cdot;c_0,c_2)$ has a root near $e^{i\theta_\tau}$, implicitly implying that one of the field directions $\{\pm u,\pm v\}$ is aligned with the tangent direction $\tau$.  Since~\eqref{eq:complex_poly} has no odd-degree terms, this term has no dependence on the sign of $\tau$, as desired.
\item \textsc{Smoothness:}  The second optimization term is a Dirichlet energy measuring the smoothness of the functions $c_0(\vec{x})$ and $c_2(\vec{x})$ as a function of the location $\vec{x}$ in the image.  Smoothly-varying $(c_0,c_2)$ pairs imply a smooth set of frame directions.  We use $\lambda=50$ in all our experiments; while the method is fairly stable to the choice of $\lambda$, larger values may be desirable for particularly noisy inputs.
\item \textsc{Regularization:}  Away from junctions, there is only one prominent direction in the image.  To prevent the frame field from collapsing into a line field, the regularization term expresses a slight preference for the field to be aligned with $\tau^\perp=g$ in the absence of other information. 
\end{itemize}

%
%The first term enforced alignment of the frame field with the corresponding tangent directions. The second one is a Dirichlet energy for the complex polynomial coefficients, enforcing field smoothness. Finally, the third term is a regularizer that prevents the frame field from collapsing into a line field, when the two direction align. In all our experiments, we use a small regularization weight ($\mu=0.01$). For all the results in the paper, we have used the same smoothness weight ($\lambda = 50$), but it may be adjusted for noisier inputs.
%
To improve results by attenuating the influence of noisy directions near junctions, we weigh the smoothness term by $\frac{w(\vec{x})}{\max_{\vec{x}}(w(\vec{x}))}$ and alignment term by $1-\frac{w(\vec{x})}{\max_{\vec{x}}(w(\vec{x}))}$, where
\begin{align}
w(\vec{x}) \eqdef \left|\frac{\langle \tau^2(\vec{x}) \rangle}{\|\langle \tau^2(\vec{x}) \rangle\|} - \tau^2(\vec{x})\right| \\
\langle \tau^2(z) \rangle \eqdef \int_{B(\vec{x})}{\tau^2(\vec{x}') d\vec{x}'}.
\end{align}
Here, $B(\vec{x})$ denotes a small (one-pixel) neighborhood of $\vec{x}$; in practice, we approximate this integral by averaging the neighboring values of $\tau^2$ adjacent to the pixel centered at $\vec{x}$.

\paragraph*{Optimization.}  We apply standard finite-difference discretization to evaluate the objective function~\eqref{eq:variational} on a pixel grid.  %In our implementation, we set $B(z)$ to 1-pixel neighborhood of the pixel at $z$. 
The end result is a quadratic, unconstrained optimization problem for a $(c_0,c_2)$ pair per pixel.  We use ``natural'' (Neumann) boundary conditions, essentially evaluating the gradient-based smoothness term only on pairs of adjacent pixels that are both included as degrees of freedom in the numerical problem.

We use the L-BFGS algorithm for optimization \cite{Nocedal2006}, with a history of 6 iterates for the quasi-Newton Hessian approximation.  Our code uses the LBFGS++ implementation described by Qiu et al.~\shortcite{LibBFGS}.  We start from an initial guess of an axis-aligned cross field.  The quadratic nature of our optimization problem would allow for more specialized techniques, e.g.\ preconditioned conjugate gradients, but as frame field computation is not currently the efficiency bottleneck of our algorithm, we leave tuning of this step to future work.

In the end, we only require frame directions on the image pixels corresponding to strokes that we will trace.  Hence, to improve optimization efficiency and to improve junction resolution even with acute angles, we take inspiration from \emph{narrow band} level set methods \cite{ADALSTEINSSON1995269} and only include variables corresponding to pixels in $I$; that is, pixels corresponding to white areas in the input image are ignored.  This greatly reduces the number of variables, yielding a significant boost in performance.

The optimization yields two scalar fields, $c_0, c_2$. At every dark pixel $i\in I$, we then use~\eqref{eq:coefftoframe} to recover the frame field directions $\{\pm u,\pm v\}$. %we then solve a biquadratic complex equation \ref{eq:complex_poly}, with the two roots - the directions of the frame field, $\pm u_i$ and $\pm v_i$.

% !TEX root = ../polyvectorization.tex

\begin{figure}
	\includegraphics[width=\linewidth]{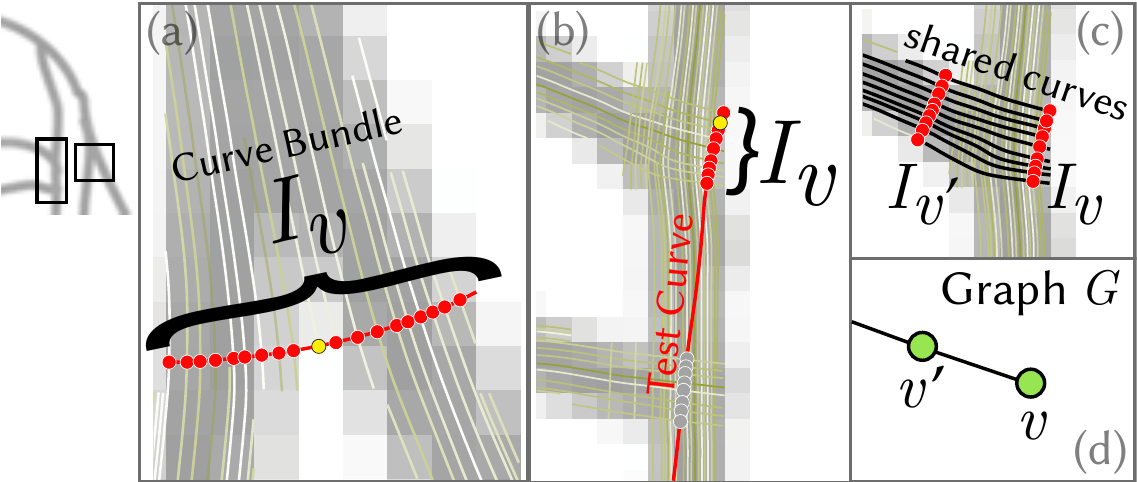}
	\caption{Grouping curves into bundles \revisedtext{and constructing graph $G$. (a,b) We trace a test curve (red) from each seed point (yellow)}, recording its intersection points with the curves \revisedtext{(red and gray}). \revisedtext{Starting with a seed point, } we then group the points that are adjacent along the test curve and closer than a pixel distance apart\revisedtext{, forming \textit{a curve bundle} $I_v$ (red). (c,d) We then associate each curve bundle with a vertex in the graph $G$, connecting vertices if they have at least one pair of intersection points adjacent along their shared curve.}}
	\label{fig:curve_bundle}
\end{figure}

\begin{figure*}
	\includegraphics[width=\linewidth]{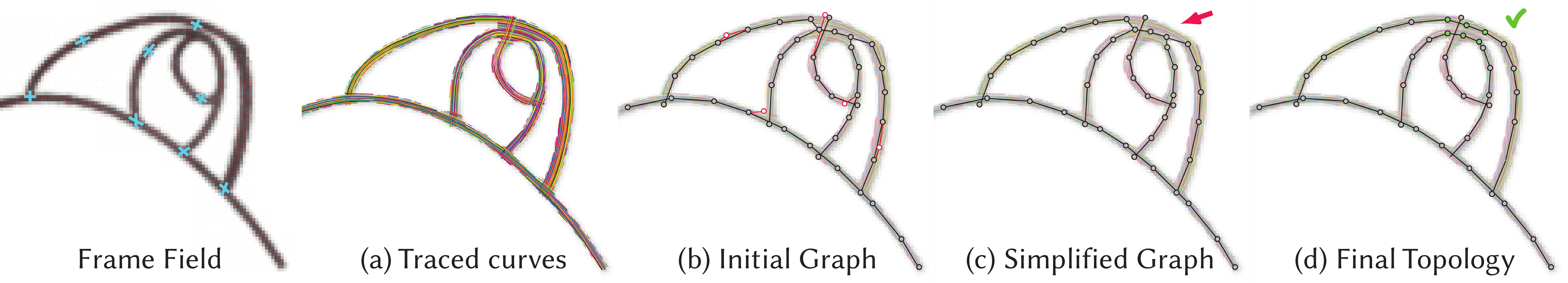}
	\caption{Stages of extracting drawing topology. Starting with the frame field (left, zoom in for complete frame field), we first trace curves passing through each dark pixel (a). We then locally group curves into curve bundles, and use adjacency along curves to form the initial graph (b). We then perform topological simplification (c), and finally, disambiguate parallel strokes, resulting in the final graph (d).}
	\label{fig:graph}
\end{figure*}

\section{Extracting Drawing Topology}
\label{sec:topology}

The next step of our algorithm extracts the topology of the drawing from the computed frame field. The key requirement is not only to extract the correct topology, but also to create a structure that allows for subsequent vectorization aligned with the frame field.

Starting from each dark pixel, we trace the frame field (Fig.~\ref{fig:graph}(a)); these curves are grouped locally into \textit{curve bundles}. Each curve bundle is associated to a vertex in a topological graph $G=(V,E)$, whose adjacency is determined by shared curve segments between different bundles (Fig.~\ref{fig:graph}(b)).
%Associating each curve bundle with a graph vertex and connecting vertices based on natural adjacency of curves bundles along the shared curves, we form a topological graph $G = (V,E)$ (Fig.~\ref{fig:graph} (b)). 
After topological simplification  (Fig.~\ref{fig:graph}(c)) and disambiguating parallel strokes, this graph has the topology of the line drawing (Fig.~\ref{fig:graph}(d)) and allows for vectorization by following shared curves connecting each pair of curve bundles. 

\subsection{Initial Graph Construction}

\revisedtext{Away from junctions, the largest root of the frame field reliably is aligned with the curve tangent. } Therefore, at every dark pixel $\vec{x}$, we choose the frame field root with the maximum magnitude; without loss of generality, we will label it $u(\vec{x})$. We then trace the frame field starting from $\vec{x}$ in both directions $\pm u$, using simple Euler's integration method with a step size $h=0.1$, where each pixel is considered to have width 1 (Fig.~\ref{fig:graph}(a)). We stop tracing as soon as a curve leaves the narrow band \revisedtext{or comes indistinguishably close (within $0.01$ distance in our implementation) to a curve with the same tangent. The latter condition is designed partly to prevent closed curves in the drawing from being traced over multiple times, and partly for efficiency reasons to avoid overtracing.}

The integration step yields more curves than will be present in the final traced image.  Hence, once all curves are traced, we \revisedtext{split up} curves into groups corresponding to strokes in the drawing. Since curves may naturally continue past acute \revisedtext{junctions or \textit{Y-junctions}---valence-3 junctions with three distinct directions---}we create a topological graph by grouping curves locally \revisedtext{(Fig.~\ref{fig:curve_bundle})}. Our goal is to group corresponding curves along the width of the stroke, perpendicular to the centerline; each group forms a vertex of the graph. %Clearly, we would like to % <-- why is this clear?
\revisedtext{We} only group curves corresponding to the same direction of the frame field, thus separating intersecting strokes.

 Starting from each curve endpoint (\textit{seed}), at each of the 8 neighboring pixels, we select a matching direction of the frame field using the standard least-angle matching criterion \cite{Diamanti2015}. We then trace a curve perpendicular to this local direction field, extending the field \revisedtext{each time we move to a neighboring pixel by the same procedure: We look at the new 8-pixel neighborhood and match the frame field directions}. Once the \revisedtext{orthogonal} test curve is traced, we find its intersection points with the curves with the matching direction. We then \revisedtext{form} the group of intersection points $I_v$\revisedtext{, the curve bundle associated }with a graph vertex $v\in V$\revisedtext{,} in the following way (Fig.~\ref{fig:curve_bundle}): \revisedtext{Starting with the seed point, we group adjacent intersection points if they are less than one pixel apart.} This strategy \revisedtext{effectively groups curves along the width of the stroke without relying on an estimate of the stroke width, based only on the simpler assumption that different parallel strokes are separated by at least one pixel (Fig.~\ref{fig:curve_bundle}(a,b)).} We then add edges between vertices that have at least one pair of intersection points adjacent along their shared curve \revisedtext{(Fig.~\ref{fig:curve_bundle}(c,d))}.
 \revisedtext{To efficiently test for intersections, in our implementation we cache segments overlapping with each pixel; we then test for intersections only with the segments in the pixels overlapping with the test curve.}

Using this graph, one can define \revisedtext{an ``\textit{induced}''} vectorization of the drawing with the same topology as the graph. Namely, each edge of the graph defines a set of curve segments connecting adjacent curve bundles. By choosing one of those segments per edge and connecting, if necessary, the ends of those segments with straight lines, we obtain a vectorization with the topology of the graph (Fig.~\ref{fig:vectorization}). This vectorization is not unique, since edges are typically associated to more than one shared curve (Fig.~\ref{fig:vectorization}(b)); we discuss the vectorization process further in Section \ref{sec:vectorization}.

\subsection{Topology Simplification}
\label{sec:topology_simplification}

\setlength{\columnsep}{0.03in}
\begin{wrapfigure}[6]{l}{0.35\linewidth}
	\vspace{-0.1in}
	\includegraphics[width=\linewidth]{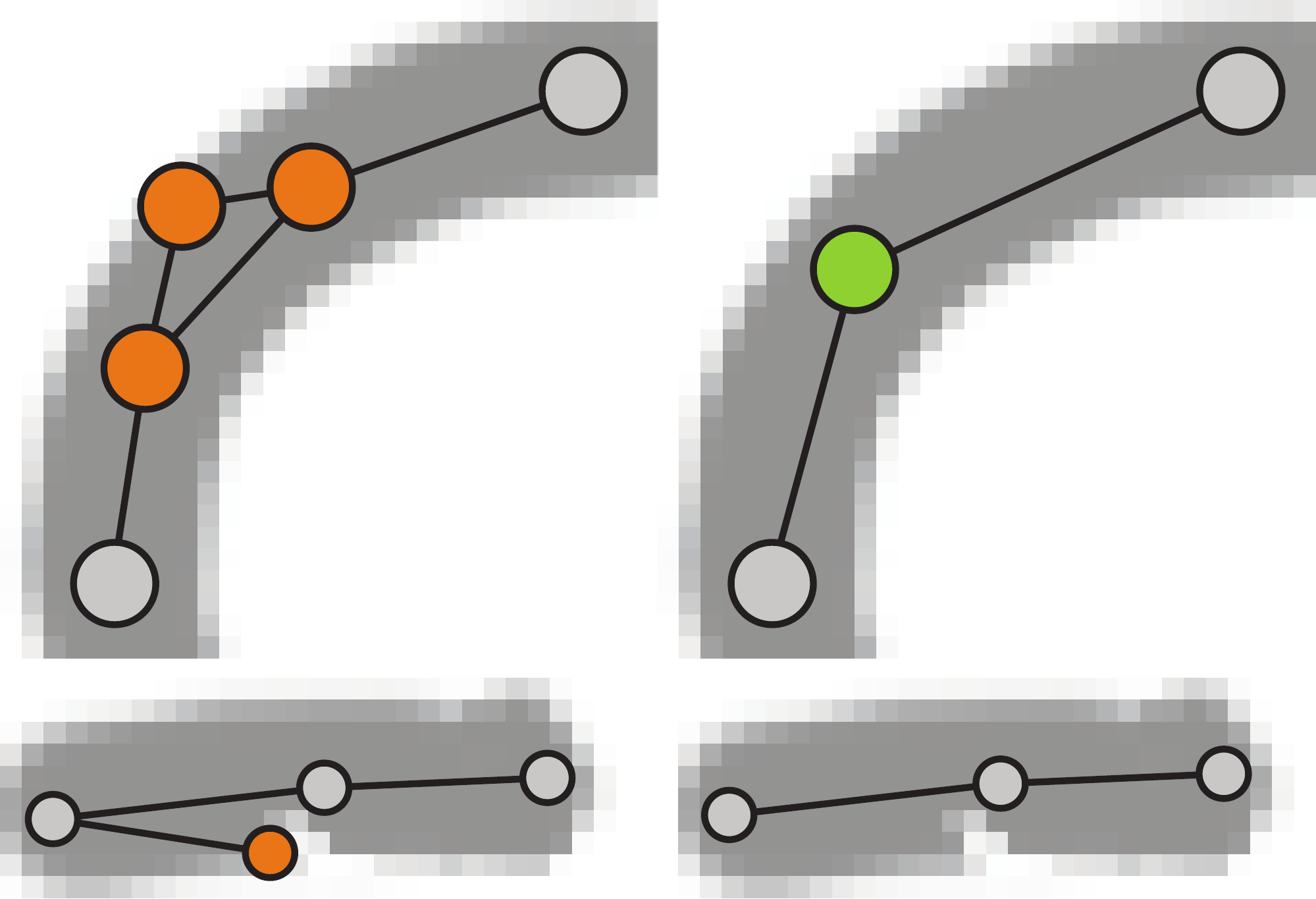}
\end{wrapfigure}

An uneven narrow band boundary can affect the topology of the graph. In particular, it might introduce extraneous loops and branches (\revisedtext{inset,} Fig.~\ref{fig:graph}(b), red). To account for this, we perform the topology simplification procedure below (\revisedtext{inset, Fig.~\ref{fig:graph}(b)-(c)}). 

 Since we expect each loop to correspond to a hole in the narrow band, we contract each loop if its induced vectorization contains fewer than $n_\mathrm{hole}$ white pixels \revisedtext{(inset, top)}. To distinguish true topological \revisedtext{\textit{branches}---valence-two paths ending with a leaf node---}from extraneous ones, we use the following heuristic. For each vertex with valence greater than two, we repeatedly choose the shortest \revisedtext{branch and prune it if its length outside the strokes formed by the other branches is too short (inset, bottom). In our implementation, for a given branch, we use a quarter of its full length, or a pixel, whichever is greater, as such threshold}. To perform this test, we define stroke width at each vertex as the maximal distance between the intersection points of its curve bundle. Then, for each vertex of the branch we find the closest vertex not belonging to the branch and test if Euclidean distance between those is within sum of their stroke radii.

\subsection{Disambiguating parallel strokes}

\setlength{\columnsep}{0.03in}
\begin{wrapfigure}[7]{l}{0.25\linewidth}
	\vspace{-0.1in}
	\includegraphics[width=\linewidth]{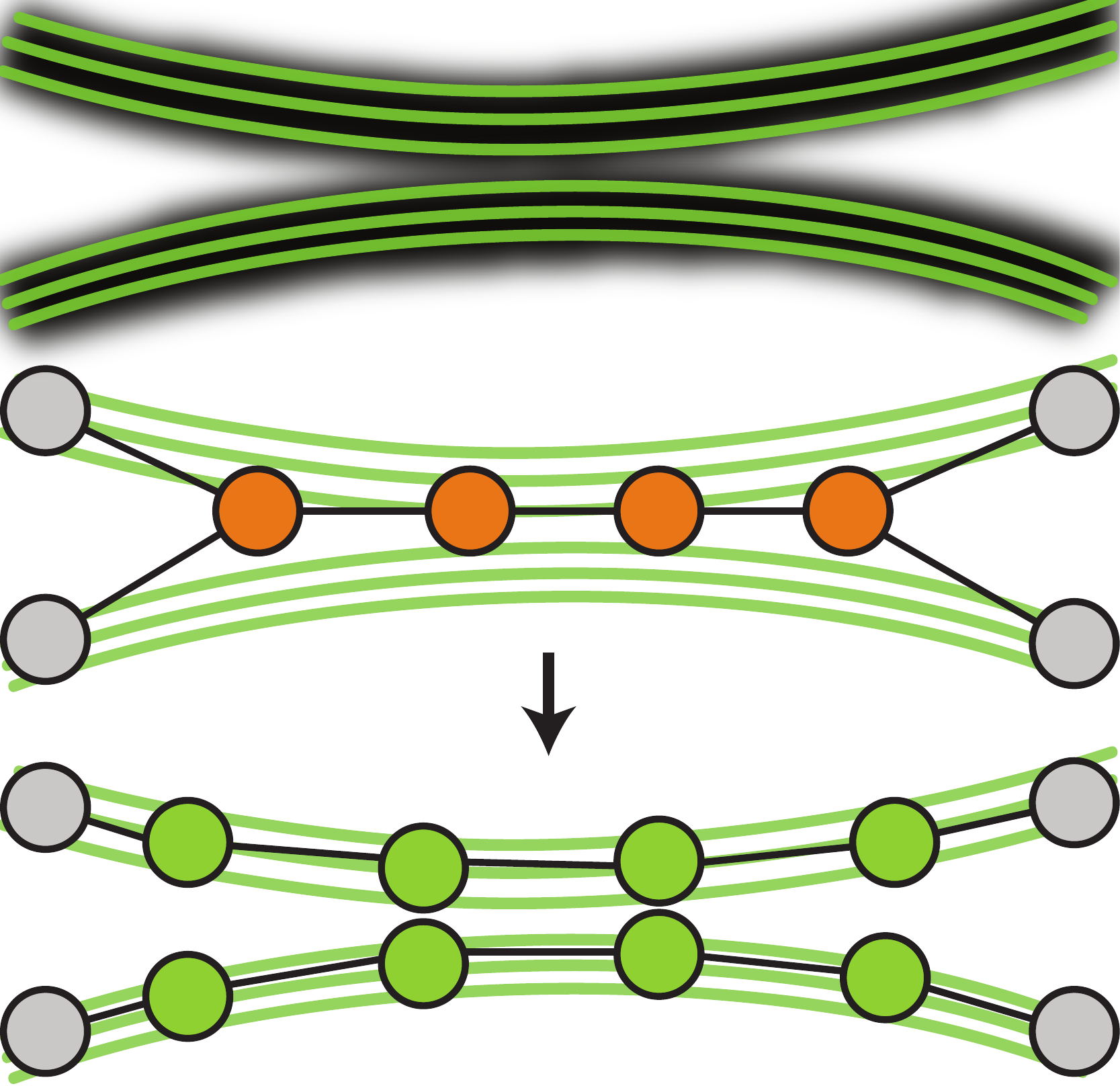}
\end{wrapfigure}

In our final stage of topology extraction, we separate parallel strokes that are merged due to a connected narrow band segment (\revisedtext{inset and} Fig.~\ref{fig:graph}(c)-(d)). \revisedtext{This happens when two different but nearly parallel strokes touch or overlap (inset, top): In the overlap, the traced curves of the upper stroke will be naturally grouped with the traced curves of the lower stroke, forming the orange vertices. To resolve this}, we first find paths of valence-2 vertices connecting pairs of vertices with valence 3 \revisedtext{(inset, orange)}.
Edges along these paths are split into two by duplicating their vertices; 
this procedure ``unzips'' the path connecting the degree-3 vertices (\revisedtext{inset and} Fig.~\ref{fig:graph}(d), green vertices). 
%
%We then connect the vertex chains to the vertices adjacent to the split set using the traced curves in the curve bundles of those vertices. 
The remaining edges at the degree-3 vertices that were not unzipped are assigned \revisedtext{to} new neighbors based on the connectivity of the underlying curve bundle. 

\setlength{\columnsep}{0.03in}
\begin{wrapfigure}[5]{l}{0.1\linewidth}
	\vspace{-0.1in}
	\includegraphics[width=\linewidth]{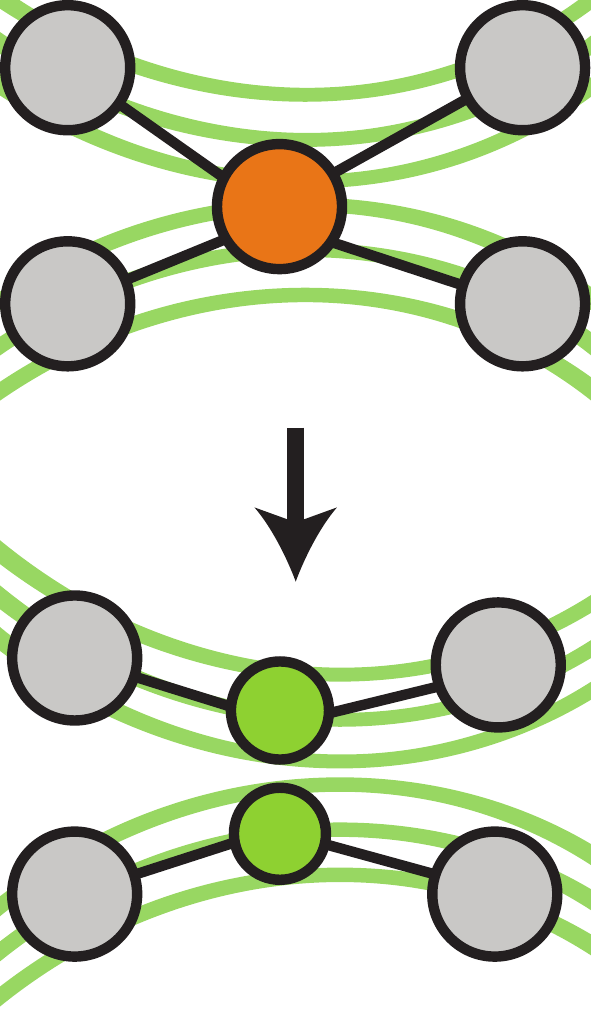}
\end{wrapfigure}

At X-junctions, two strokes intersect but they do not share a vertex in our graph construction.  Hence, vertices with valence 4 or higher are extremely rare. We split these vertices using the same unzipping technique, effectively treating them as a pair of degree-3 vertices connected by an edge of length zero.%, treating this vertex as a degenerate split set.

%In our final stage of topology extraction, we separate parallel strokes that are merged due to a connected narrow band segment (Fig.~\ref{fig:graph}(c)-(d)). To do this, we find the \textit{split set}, graph edges connecting two vertices with valence 3, and split them into two edges each, duplicating their adjacent vertices. Thus, each split set generates two degree-2 vertex chains (Fig.~\ref{fig:graph}(d), green vertices). We then connect the vertex chains to the vertices adjacent to the split set using the traced curves in the curve bundles of those vertices. Since at X-junctions the two intersecting strokes are separated in the graph by construction, vertices with valence 4 or higher are extremely rare. We split vertices of degree 4 in the same fashion, treating this vertex as a degenerate split set.

\subsection{Treating Frame Field Singularities}
In contrast to %frame field and quad meshing literature 
\revisedtext{frame fields applied to quad meshing}
\cite{Diamanti2015}, singularities in our frame field do not have meaningful interpretation; they usually are artifacts due to noise. Our insight is since singularities happen in the areas with significant noise, we can eliminate most of the singular points by relaxing the alignment term in those areas. Therefore, after the frame field optimization, we find singular pixels, set their alignment weight to zero, and re-run the optimization \eqref{eq:variational}. We repeat this process, \revisedtext{each time updating the alignment weights,} until no more singularities can be resolved this way. \revisedtext{Since each time we either reduce the number of non-zero alignment weights or stop, and since a frame field with no alignment term has no singularities, the process necessarily terminates. In practice, however, the alignment only needs to be relaxed for a small number of pixels, typically less than 1\% of dark pixels. }

\begin{figure}
	\includegraphics[width=\linewidth]{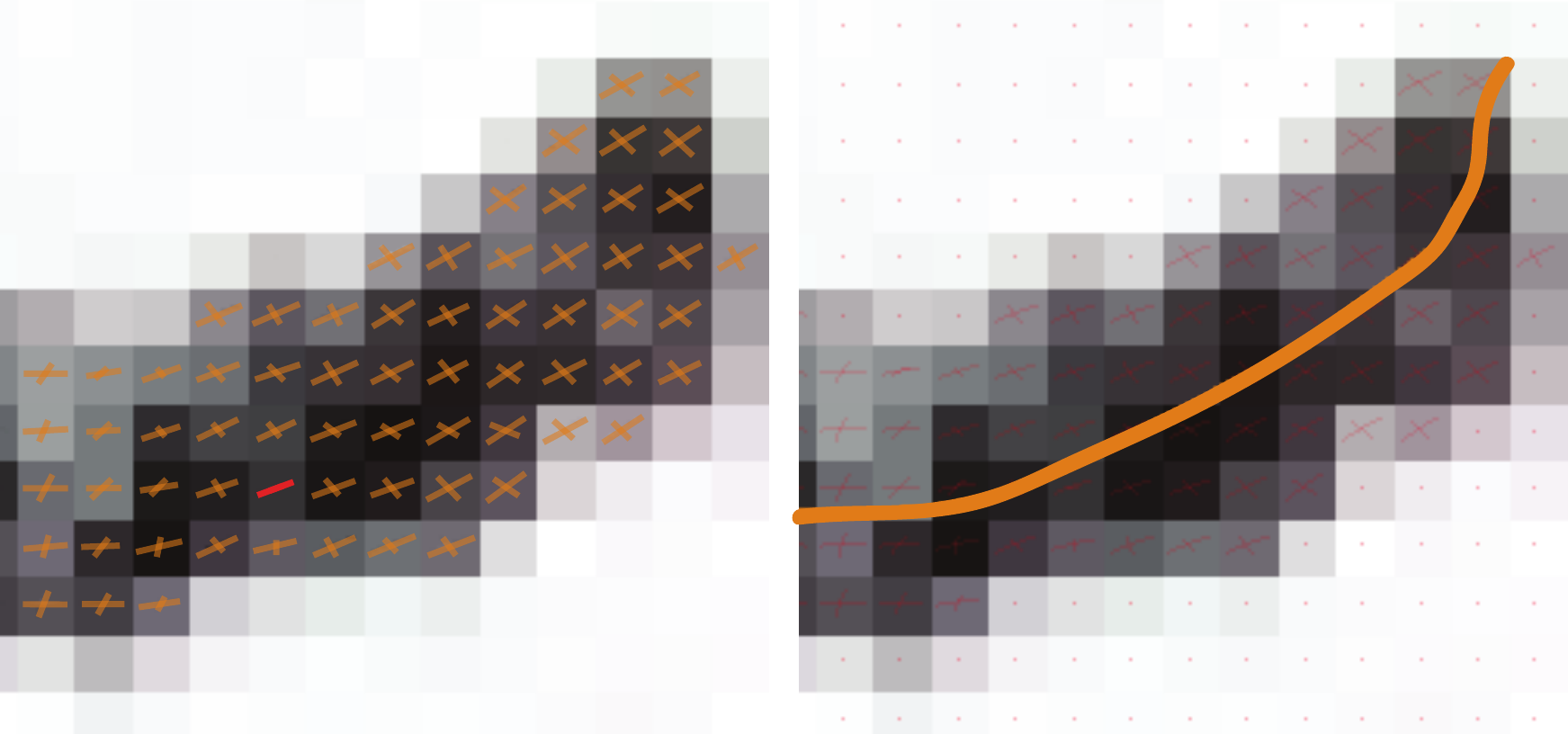}
	\caption{\minorrev{(left) A typical singularity (highlighted in red) occurs at a point where the two frame directions coincide. (right) Final vectorization.}}
	\label{fig:singularity}
\end{figure}

We address the remaining singularities (typically fewer than five singular pixels per image, \minorrev{Fig. \ref{fig:singularity}}) using a simple heuristic. First, we stop tracing curves at singular pixels. This ensures that no vertices in the topological graph with inconsistently matched frame field directions are connected. However, this may introduce a gap in a stroke in the final vectorization. To address this, we mark leaf vertices next to a singularity in the topological graph, and greedily connect each to the closest non-adjacent vertex.

\section{Vectorization}
\label{sec:vectorization}
We use the extracted topological graph $G$ to create the final vectorization. The key idea is to extract a vectorization that follows the traced curves as much as possible while having the topology of the graph. Additionally, we would like the vectorized curves close to the centerline of the stroke.   %Given a topological graph, we can directly generate a vectorization with the same topology as the graph in the following way. 

We initialize our procedure for embedding the topological graph by embedding vertices $v$ whose degree does not equal two; these vertices correspond to isolated stroke endpoints as well as points where curve segments join together tangentially. 
Recall that $v$ in the topological graph represents a \emph{bundle} $I_v$ of intersection points between the traced curves and an orthogonal test curve (red points in Figure~\ref{fig:curve_bundle}). \revisedtext{Thus, a natural choice of embedding for each vertex $v$ is one of the points in $I_v$.}
With this construction in mind, we approximate the drawing centerline at $v$ as the barycenter $b_v$ of $I_v$ (Fig.~\ref{fig:tracing_cost}, green points); $b_v$ becomes the assigned position for $v$ in our embedding (Fig.~\ref{fig:vectorization}(b)).

What remains is to embed the degree-2 vertices and the edges of the graph. Our objective in this step is to embed edges as curves that approximately follow the traced curves without diverging too far from the stroke centerline. \revisedtext{Each individual traced curve, even if it started at the stroke center, might drift away from the centerline. To account for that, we select a curve on a per-edge basis}; our vectorization can ``hop'' from one traced curve to another at the vertices of the topological graph, in which case the two \revisedtext{curve segments} are connected using a straight line segment.  The end result is a tracing that is composed piecewise of traced curves connected with short ligaments that subsequently can be smoothed. \revisedtext{The details of this procedure are outlined below. }

\subsection{\revisedtext{Auxiliary Graph}}

\begin{figure}
	\includegraphics[width=\linewidth]{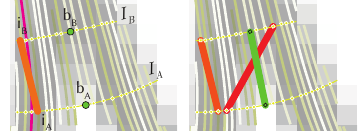}
	\caption{A few vertices and edges of the auxiliary graph. Left: a cost of an edge (orange) between two vertices in the auxiliary graph is computed as a \revisedtext{weighted sum of distances between $i_{A,B}$ and the barycenters $b_{A,B}$ ($E_{\mathrm{centering}}$)} and a sum of distances to the closest shared curve (magenta, $E_{\mathrm{connections}}$). This penalizes edges corresponding to vectorizations far from center (orange) or vectorizations deviating from the traced curves (red). Using stroke width, computed in \S\ref{sec:topology_simplification}, in the centering term relaxes centering around Y-junctions, where the stroke width is locally high\revisedtext{.}}
	\label{fig:tracing_cost}
\end{figure}

\begin{figure}
	\includegraphics[width=\linewidth]{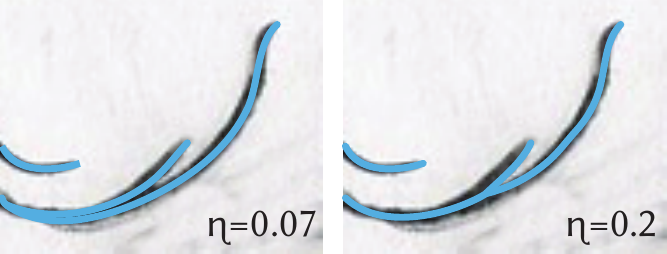}
	\caption{The centering weight $\eta$ controls the locations of Y-junctions.}
	\label{fig:changing_eta}
\end{figure}

\revisedtext{W}e cast the remaining embedding computation as a shortest-path problem over an \textit{auxiliary graph} $G^{Aux} = (V_G^{Aux}, E_G^{Aux})$ constructed as follows:
\begin{itemize}[leftmargin=*]
\item \textsc{Vertices} ($V_G^{Aux}$):  The vertices of the auxiliary graph are defined as the union $\cup_{v\in V}I_v$, corresponding to the set of \emph{all} intersection points between the traced curves and the test curves (yellow points in Fig.~\ref{fig:tracing_cost}).
\item \textsc{Edges} ($E_G^{Aux}$):  Recall the vertices of $G^{Aux}$ are clustered into sets $I_v$.  For two vertices $v_1,v_2$ connected by an edge $(v_1,v_2)\in E$ in the topological graph (rather than $G^{Aux}$), we insert a bipartite graph of edges connecting all vertices in $I_{v_1}$ to all vertices in $I_{v_2}$. Symbolically, we can write:
$$
E_G^{Aux} = \{
(p,q) : p\in I_{v_1}, q\in I_{v_2}, (v_1,v_2)\in E
\}.
$$

\revisedtext{Intuitively, by following an edge in the auxiliary graph, we connect the two intersection points with a segment of some traced curve shared by their curve bundles and, possibly, a straight line segment.}

\item \textsc{Edge weights} ($w_e$):  \revisedtext{The edge weight is designed so that \minorrev{the} shortest path on the graph produces a curve that is smooth and centered. Thus, t}he weight of an edge $e$ is a weighted sum of two terms: $$w_e = E_{\mathrm{connections}} + \eta E_{\mathrm{centering}}.$$  
Roughly, the first term penalizes hopping from one curve to another when they are far away, \revisedtext{in some sense favoring smoother connections;} and the second term is designed to penalize connecting pairs of vertices that are far from the centerline \revisedtext{(Figure~\ref{fig:tracing_cost}). 

More concretely, the first term, $E_{\mathrm{connections}}$, is computed as a sum of distances to the closest shared curve (for an orange edge on Fig.~\ref{fig:tracing_cost} it is the magenta curve): 	$$
	E_{\mathrm{connections}} = \min_{r_{1,2} \in I_{v_{1,2}}, c(r_1)=c(r_2)} {[\|r_1-p\| + \|r_2-q\|]},
	$$
	where $c(r)$ is the traced curve containing the intersection point $r$ of some curve bundle. 
	The centering term penalizes the distance from a vertex to the corresponding barycenter, i.e.\ for an edge $e = (p, q) \in E^{Aux}_G$ from bundles $I_{v_1}$ and $I_{v_2}$ with stroke widths $w_{v_1},w_{v_2}$ respectively, 
	$$E_{\mathrm{centering}} = \frac{\|p - b_{v_1}\|+\|q - b_{v_2}\|}{w_{v_1} + w_{v_2}}.$$} The centering weight $\eta$ affects the exact locations of Y-junctions in ambiguous areas (Fig.~\ref{fig:changing_eta}); we use $\eta=0.07$ in all our experiments.
%For each edge connecting $i_{v_1}\in I_{v_1}$ to $i_{v_2}\in I_{v_2}$, $E_{\mathrm{connections}}$ is a sum of Euclidean distances between each intersection point and the closest intersection point on a shared curve passing through $I_A$ and $I_B$ (Fig.~\ref{fig:tracing_cost}). Thus, this term measures the length of the straight-line connections and is 0 if the two intersection points belong to the same curve. $E_{\mathrm{centering}}$ is a sum of Euclidean distances between $i_A, i_B$ and barycenters of $I_A, I_B$ (Fig.~\ref{fig:tracing_cost}), divided by their respective stroke widths estimates, computed in Sec. \ref{sec:topology_simplification}. This allows to reduce the influence of the centering term around Y-junctions, where the stroke width is locally high, to prefer a smooth transition of the curves based on the frame field to a behavior similar to medial axis (similar to Fig.~\ref{fig:changing_eta}(a) vs. (b)).
\end{itemize}

\subsection{\revisedtext{Extracting the Vectorization}}
\label{sec:refining_vectorization}
\begin{figure}
	\includegraphics[width=\linewidth]{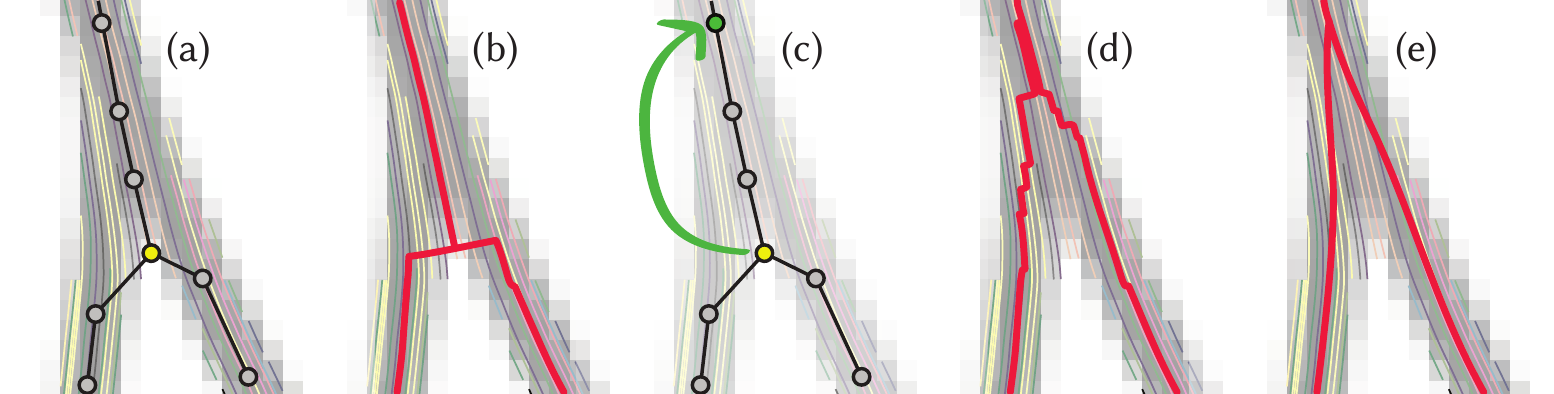}
	\caption{Generating final vectorization from the topological graph. (a) Topological graph, (b) A vectorization, (c) Optimization result, (d) Optimized vectorization, (e) Final smooth result}
	\label{fig:vectorization}
\end{figure}

As an initial embedding of the full graph \revisedtext{$G$}, we simply connect the previously-embedded vertices \revisedtext{(with degree not equal two)} using shortest paths in this weighted \revisedtext{auxiliary} graph.  This produces a vectorization with the correct topology, which is centered and follows the traced curves \revisedtext{(Fig. \ref{fig:vectorization}(a,b))}. 

After this initial pass computes an embedding, we make a second pass to refine the result.  Principally, we improve the locations of the valence-3 vertices, which can be suboptimal since they were chosen \revisedtext{independently} (Fig.~\ref{fig:vectorization}(b)). %Therefore, we further improve the vectorization by optimizing location of each degree-3 vertex as follows. 
Our procedure for moving the valence-3 vertices to improved locations is illustrated in Figure~\ref{fig:vectorization} and described below.

%Since each barycenter corresponds to a vertex in $G_A$, w
%
%We seek an assignment of a barycenter $b_v$ of a bundle $I_v$ for each degree-3 vertex to further reduce the shortest-path cost for the embedding described above.  
%
%such that the total path cost between those vertices over $G_A$ is minimized. 
\revisedtext{Valence-3 vertices typically correspond to acute junctions or Y-junctions. In this stage, we find optimal locations that provide a smooth transition between the joining curves. Thus, we attempt to further improve the \revisedtext{total shortest-path cost over the graph} while preserving topology. To do so, we allow each degree-3 vertex (inset, yellow) to snap to any barycenter along its outgoing degree-2 chains in $G$ (inset, green) and find the optimal locations for valence-3 vertices minimizing the total shortest-path cost on the auxiliary graph ($G^{Aux}$, gray vertices in the inset below).}

\setlength{\columnsep}{0.03in}
\begin{wrapfigure}[10]{l}{0.3\linewidth}
	\vspace{-0.1in}
	\includegraphics[width=\linewidth]{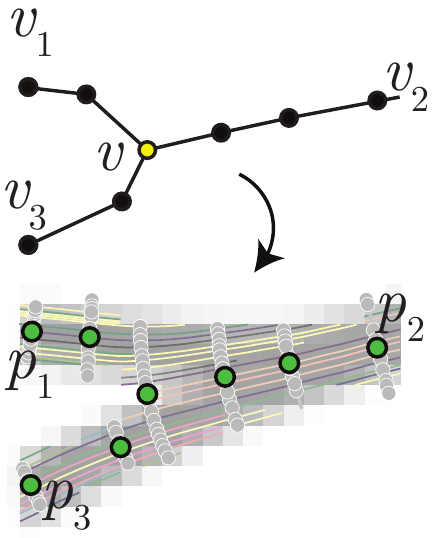}
\end{wrapfigure}

If two degree-3 vertices are connected by a chain of vertices of valence two, we fix the location of the vertex closest to the middle of the chain that was not split during the procedure described in \S\ref{sec:topology_simplification}; this avoids having to solve a global problem to place all the degree-3 vertices simultaneously and is well-justified since traditional 1-skeleton-based image vectorization methods \cite{Noris2013} perform well away from junctions. \revisedtext{After this step, every valence-3 vertex $v \in V$ is connected, via chains of degree-2 vertices, to vertices $v_1, v_2, v_3 \in V$ with fixed locations $p_1, p_2, p_3 \in V_G^{Aux}$ (inset). Denoting the set of all the vertices in the degree-2 chains connecting $v$ to $v_i$ as $V_c \subseteq V$, we restrict the set of possible embedding locations for the valence-3 vertex $v$ to the set of all curve bundle barycenters $B = \{b_{v'} | v' \in V_c\} \subseteq V_{G^{Aux}}$ (inset, green points). In particular, $p_i \in B$. We iterate over $B$ to solve a discrete problem for the embedding of $v$:}
\begin{equation}
\revisedtext{\min_{p \in B}{\sum_{i=1,2,3} d_{G^{Aux}}(p,p_i)}},
\end{equation}
\revisedtext{where $d_{G^{Aux}}(p,p_i)$ is the shortest-path distance on $G_{Aux}$.}

\begin{figure}
	\includegraphics[width=\linewidth]{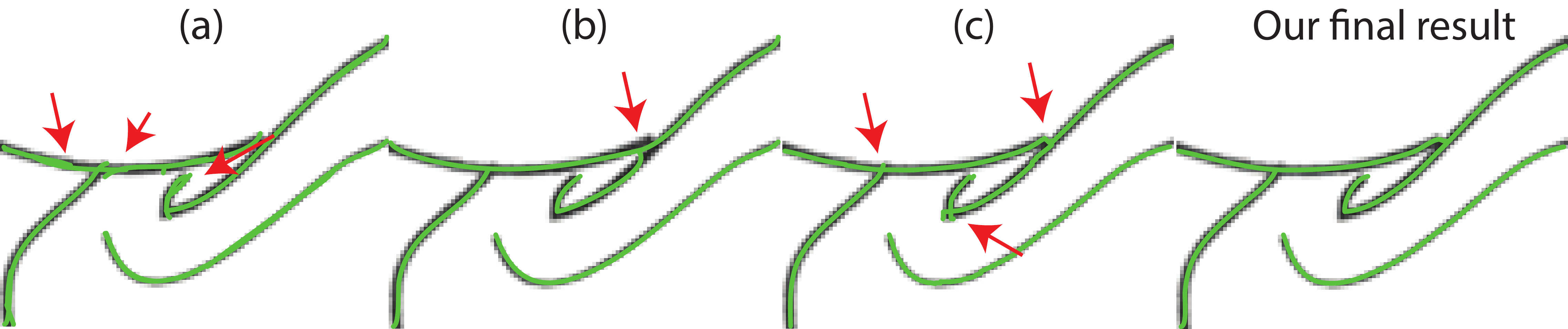}
	\caption{\revisedtext{(a) A result of our algorithm without topology simplification (Sec. \ref{sec:topology_simplification}). (b) A result without optimizing the locations of the degree-2 vertices (Sec. \ref{sec:refining_vectorization}). (c) A result without post-processing stage (end of Sec. \ref{sec:refining_vectorization}). (right) Our final result with all stages enabled.}}
	\label{fig:ablation_study}
\end{figure}

\begin{figure}
	\includegraphics[width=\linewidth]{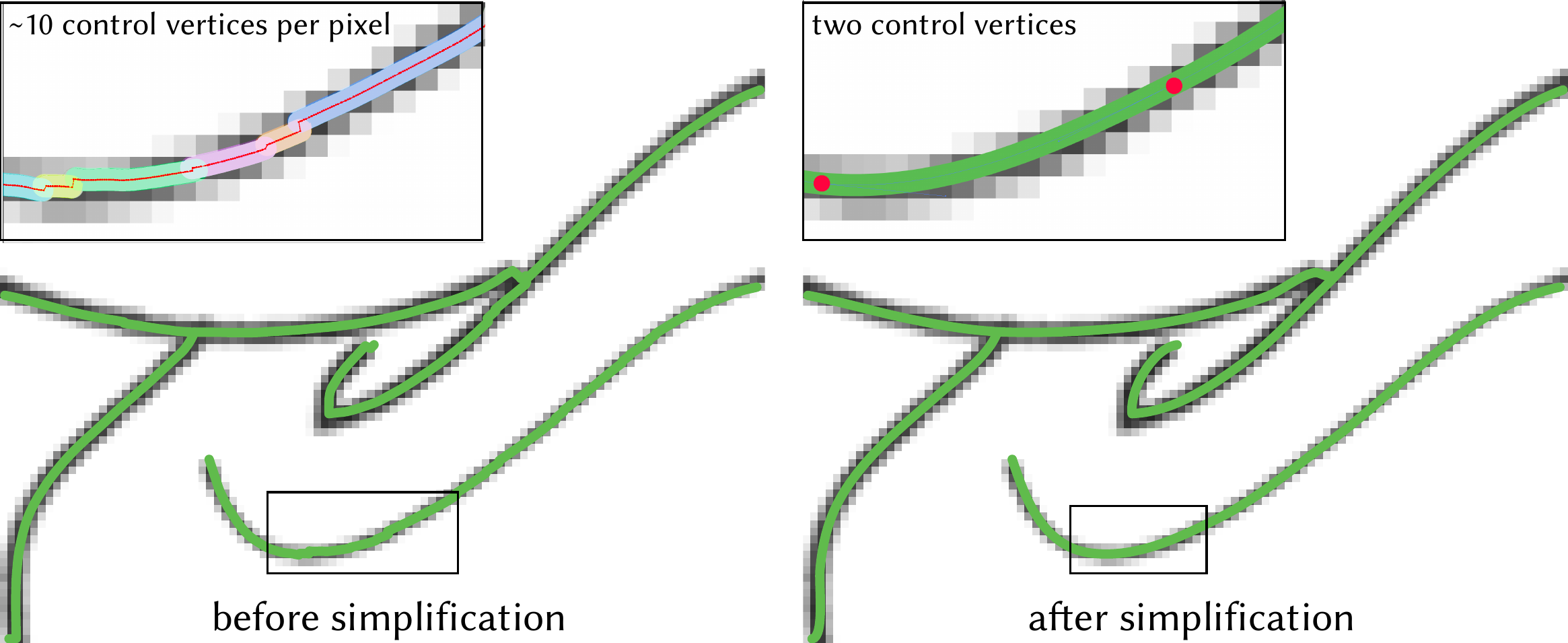}
	\caption{\revisedtext{In our implementation, we first output non-smooth curves (left) with control vertices sampled very densely (left, zoom, control vertices in red). We then use Adobe Illustrator's \minorrev{`}Simplify' feature that smooths and simplifies the curves (right). For example, after simplification the cut-out piece of curve (right, zoom) will only have two control vertices, making it easier to manipulate. }}
	\label{fig:postprocessing}
\end{figure}

\setlength{\columnsep}{0.03in}
\begin{wrapfigure}[6]{l}{0.21\linewidth}
	\vspace{-0.15in}
	\includegraphics[width=\linewidth]{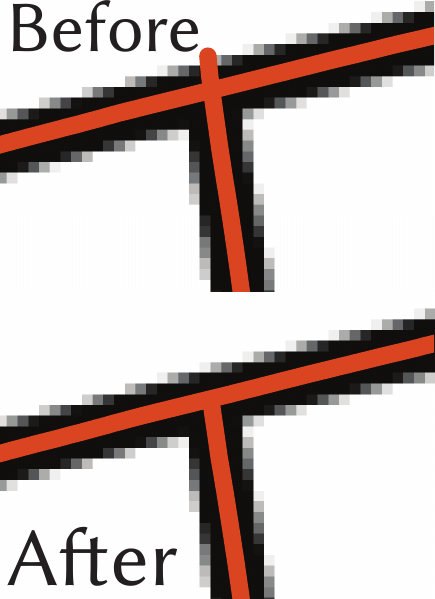}
\end{wrapfigure}

A few post-processing steps conclude our second pass. 
Since intersecting strokes are separated in the graph construction and thus are traced independently, they may continue past the points where they should meet (see inset). To prune the resulting curve fragments, we add the intersection points into the graph and repeat the branch pruning procedure (\S\ref{sec:topology_simplification}). Finally, we optionally smooth the curves using Adobe Illustrator's `Simplify Path' feature with the 95\% curve precision and zero angle threshold (Fig.~\ref{fig:vectorization}, (e) and Fig.~\ref{fig:postprocessing}). \minorrev{Alternatively, one may use the Douglas--Peucker algorithm, followed by Laplacian smoothing; this strategy produces similar results.}
% !TEX root = ../polyvectorization.tex

\section{Validation and Results}

\begin{figure*}
	\includegraphics[width=\textwidth]{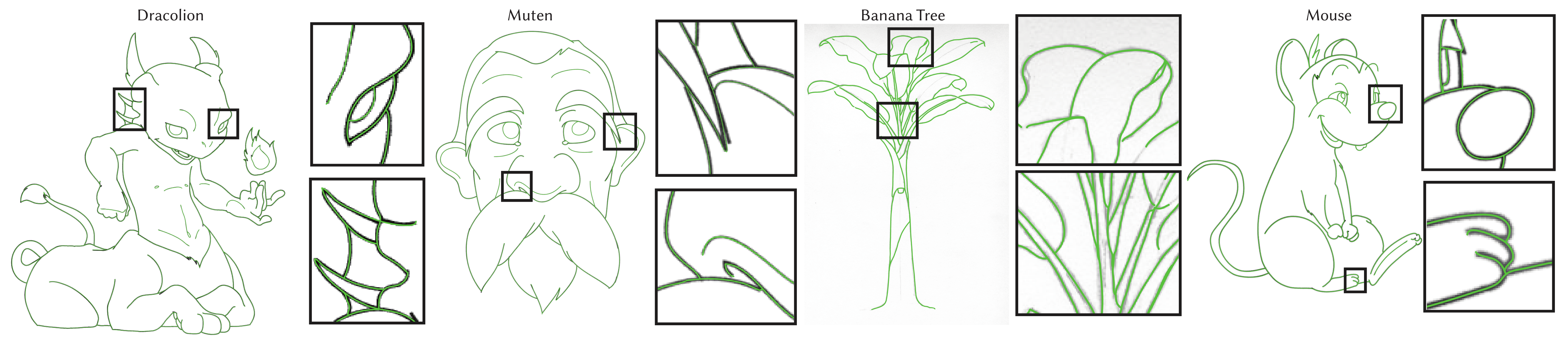}
	\caption{A gallery of additional results.}
	\label{fig:gallery}
\end{figure*}

\begin{figure*}
	\includegraphics[width=\textwidth]{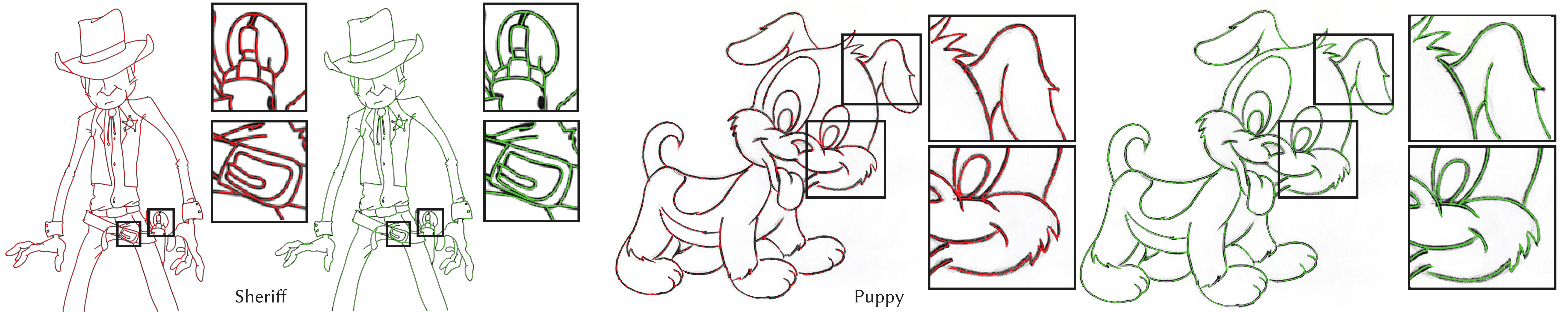}
	\caption{The method by Noris et al. \shortcite{Noris2013} (red curves) is intended to work on clean drawings only. We (green curves) obtain results of similar quality to Noris et al. on clean inputs (left, sheriff). Our method is more robust to significant noise in the drawings (right, puppy).}
	\label{fig:comparison_noris}
\end{figure*}

\begin{figure*}
	\includegraphics[width=\textwidth]{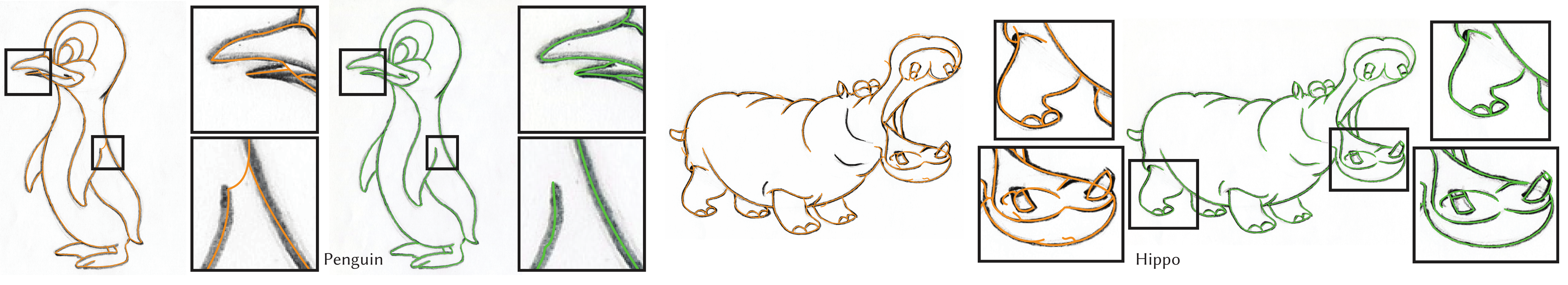}
	\caption{Our method (green curves) is aimed at truthfully vectorizing images even in the presence of noise. The method by Favreau et al. \shortcite{Favreau2016} (orange curves) is intended to work on drawings requiring significant topological simplification, and as such can be seen as complementary to our method. }
	\label{fig:comparison_favreau}
\end{figure*}

\begin{table}
	\centering
	\small
	\begin{tabular}{|l|c|c|c|c|c|}
		\hline
	& input & n. of dark & Noris & Favreau & our  \\
		& res. & pixels & et al. time & et al. time & time \\ \hline
		Muten & 1024\textsuperscript{2} & 35868 & 13s & 375s & 24s\\ \hline
		Mouse & 1024\textsuperscript{2} & 50298 & 17s & 341s & 64s\\ \hline
		Dracolion & 1024\textsuperscript{2} & 39402 & 15s & 415s & 25s\\ \hline
		Sheriff & 1024\textsuperscript{2} & 50198 & 19s & 437s & 49s\\ \hline
		Puppy & 660x624 & 29908 & 26s & 224s & 41s\\ \hline
		Hippo & 700x535 & 25114 & 24s & 120s & 43s\\ \hline
		Banana Tree & 589x865 & 18619 & 15s & 244s & 23s\\ \hline
		Penguin & 500x714 & 24134 & 23s & 181s & 56s\\ \hline
		Kitten & 700x554 & 29023 & 38s & 250s & 81s\\ \hline
		Elephant & 500x753 & 34569 & 33s & 270s & 55s\\ \hline
	\end{tabular}
	\caption{Algorithm statistics for different curve networks. }
	\label{table:time}
\end{table}

\paragraph{Qualitative Evaluation} We have automatically generated a number of vectorizations for line drawings of different style and level of noise (Figs.~\ref{fig:gallery}, \ref{fig:comparison_noris},~\ref{fig:comparison_favreau},~\ref{fig:teaser}, and ~\ref{fig:limitations} (green curves)). These included noisy, complex drawings (`Puppy,' `Elephant,' `Banana Tree'), some with varying stroke width (`Hippo,' `Penguin'), www.easy-drawings-and-sketches.com, \textcopyright Ivan Huska, as well as high-resolution clean digital images (`Sheriff,' `Dracolion,' `Muten,' `Mouse'). For all noisy images from the drawing tutorial website, to simplify line separation from the background, we automatically adjusted contrast in Adobe Photoshop. \minorrev{Alternatively, one may use an implementation of histogram equalization.}

\begin{figure}
	\includegraphics[width=0.9\linewidth]{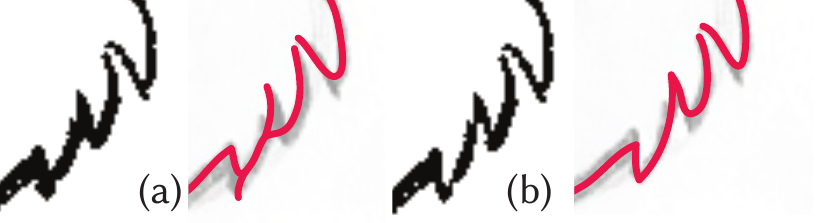}
	\caption{In the presence of noise, distinguishing junctions (a) from spikes (b) is problematic. Instead of relying on heuristics, we allow user to edit the narrow band, resulting in an arguably better interpretation (b).}
	\label{fig:user_int}
\end{figure}

The puppy example (Fig.~\ref{fig:comparison_noris}) has what Noris et al. \shortcite{Noris2013} call \textit{spikes} on the sides of the face (Fig.~\ref{fig:user_int}). In the presence of noise, distinguishing those from true junctions is problematic, and the heuristic suggested by Noris et al. \shortcite{Noris2013} breaks down. Instead of relying on similar heuristics, we allow for a simple user interaction: the user is able to edit the narrow band as a bitmap image. For this example, using a couple of brush strokes within a few seconds, the user adjusted the narrow band to achieve the desired effect (Fig.~\ref{fig:user_int}, (b)). All other input images were processed in a fully automatic way.

\begin{figure*}
	\includegraphics[width=0.9\linewidth]{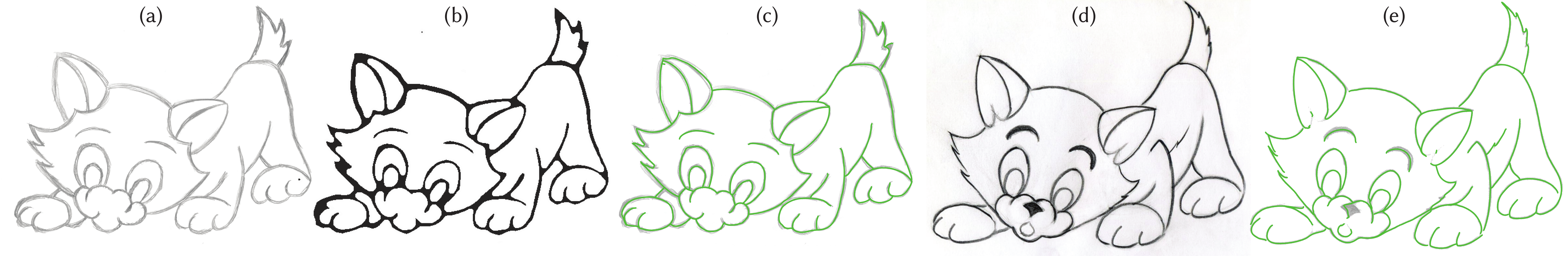}
	\caption{Our method is robust to minor input changes: \revisedtext{compare our result (c) for the input image (a), and our result (e) for a similar input image (d)}. Input image \revisedtext{(d)} is from www.easy-drawings-and-sketches.com, \textcopyright Ivan Huska; \revisedtext{the other input image (a) is from \cite{Favreau2016}}. Since our method is not aimed at drawings with fuzzy lines, we have filtered the left sketch using \cite{Bartolo2007} \revisedtext{(b)} but trace curves starting from dark pixels of the original sketch.}
	\label{fig:robustness}
\end{figure*}

\begin{figure}
	\includegraphics[width=0.5\linewidth]{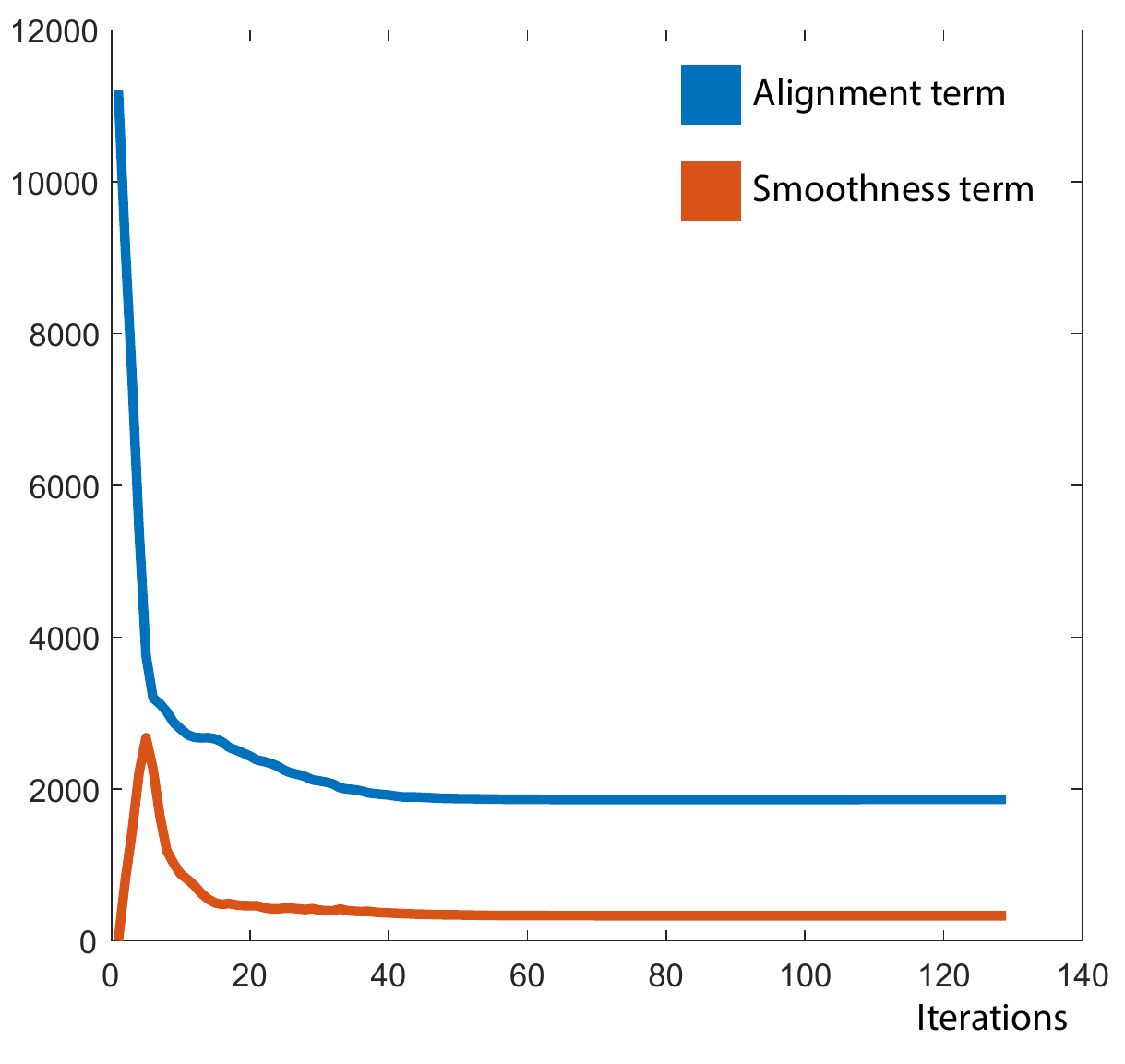}
	\caption{A plot of energy terms in Equation \ref{eq:variational} during iterations for a sample input (a cut of 'Penguin' image). The smoothness term is scaled by $\lambda$.}
\end{figure}

\begin{figure}
	\includegraphics[width=0.9\linewidth]{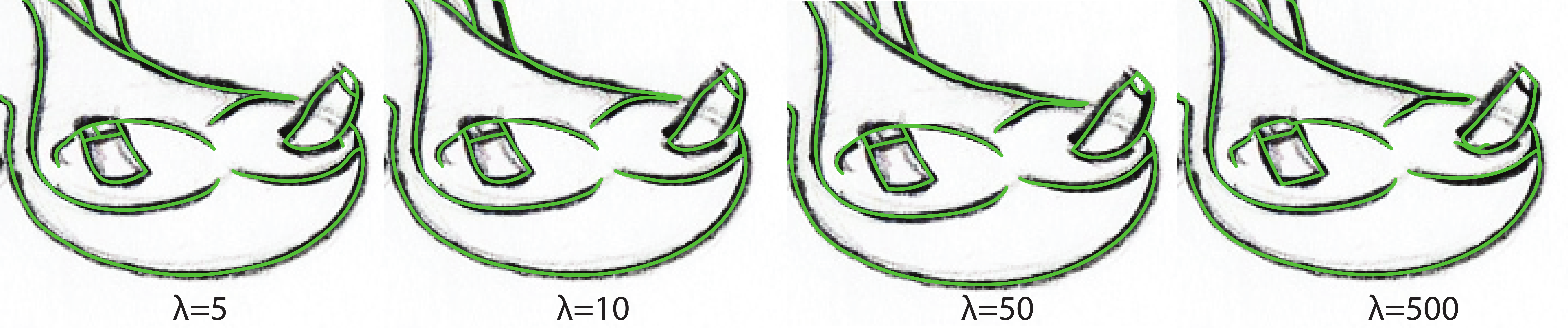}
	\caption{Our method is robust to significant changes in the frame field smoothness weight $\lambda$. Increasing the weight makes junctions sharper at a cost of losing fine details in the drawing, so higher values of $\lambda$ may be appropriate for very noisy drawings.}
	\label{fig:changing_parameters}
\end{figure}

\begin{figure}
	\includegraphics[width=0.9\linewidth]{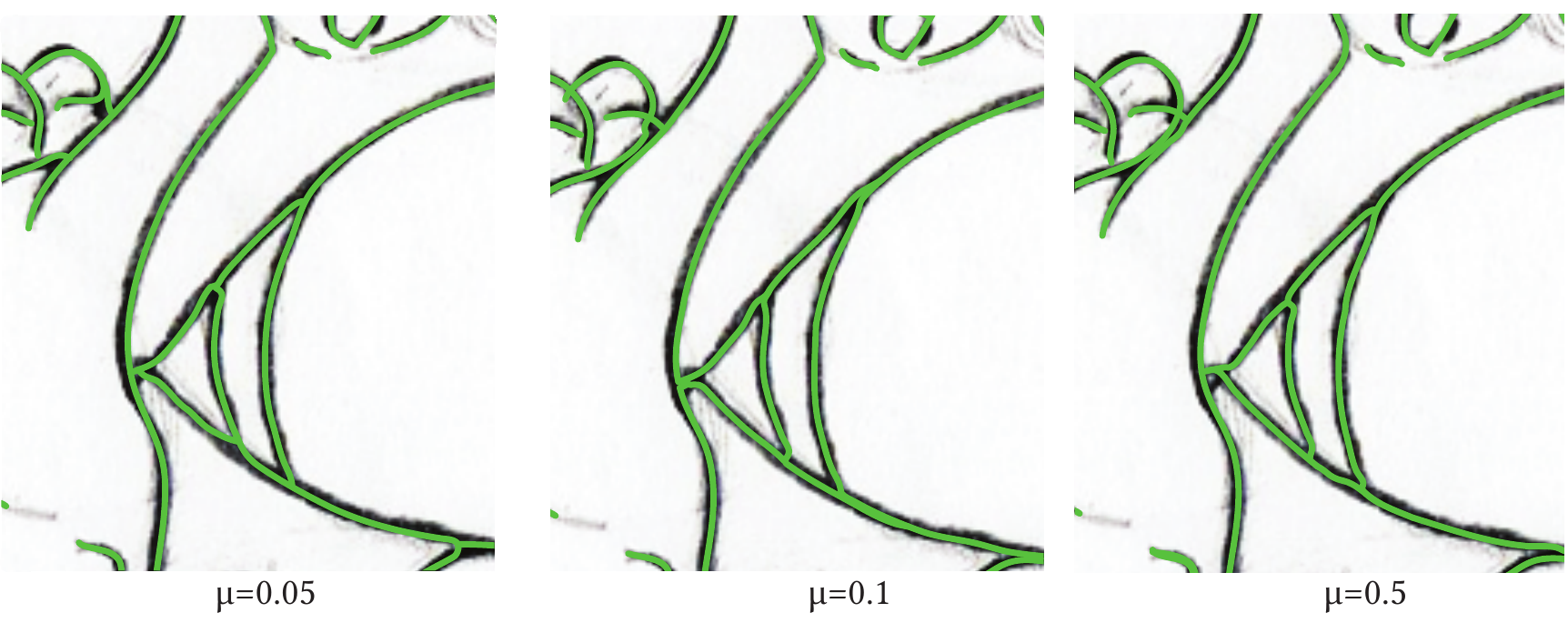}
	\caption{Our method is robust to significant changes in the frame field regularizer weight $\mu$.}
	\label{fig:changing_mu}
\end{figure}

\begin{figure}
	\includegraphics[width=0.9\linewidth]{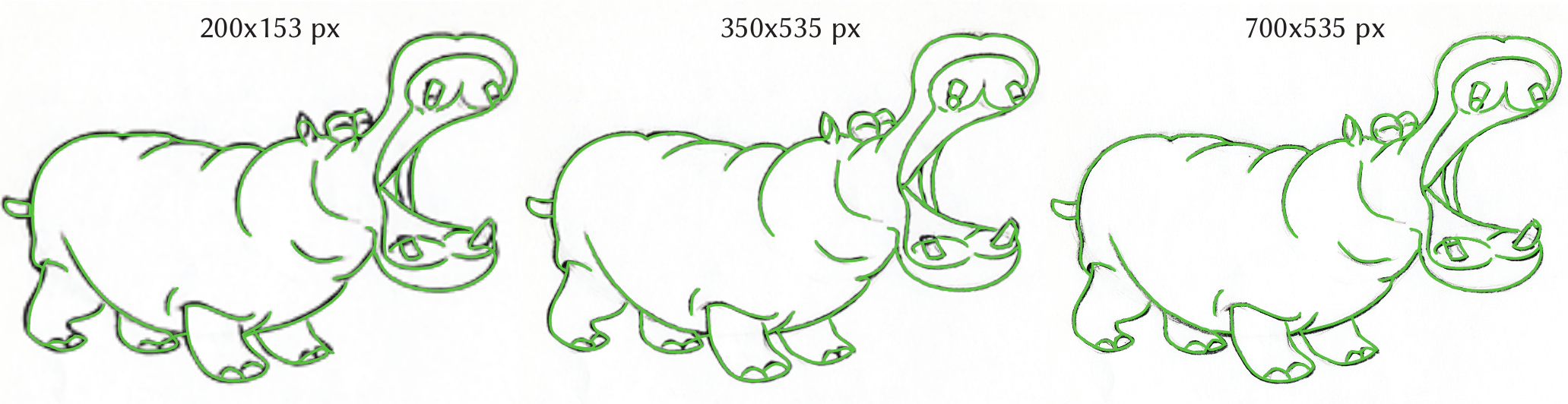}
	\caption{In general, our method is robust to significant changes in resolution. However, at low resolutions small details might be indistinguishable from noise, and thus missing in the final result.}
	\label{fig:changing_resolution}
\end{figure}

\paragraph{Comparison to Prior Art}
 We compare our method to the most relevant recent work on vectorization, described in \cite{Noris2013} and \cite{Favreau2016} (Fig.~\ref{fig:comparison_noris},\ref{fig:comparison_favreau}). %, using the implementations graciously shared by the authors of those papers.
 
 To run the method by Favreau et al.~\shortcite{Favreau2016}, we try two sets of input parameters: the default parameters in their implementation\footnote{maxNumOpenCurves = 0, minLengthOpenCurves=30, minRegionSize=7} and ones manually selected to improve results;\footnote{maxNumOpenCurves = 30, minLengthOpenCurves=5, minRegionSize=3} we keep the `fidelity-simplicity' weight at the default value of 0.5 ($\lambda$ in their formula~(2)). To run the method by Noris et al. \shortcite{Noris2013}, we try a set of parameter values, including the default parameters in their implementation,\footnote{All combinations of: Maximal Interact Distance $\in \{0.5,1.0,1.1,1.2\}$, Maximal Active Distance $\in \{0.4, 1.0, 1.1\}$, Direction Threshold $\in \{0.04, 0.05, 0.06\}$} and choose the best result. We tried both thresholding the initial images using our parameter value of $\theta_{\textrm{noise}}$, as well as not thresholding. Optionally, we additionally run a post-processing step on the results by Noris et al. \shortcite{Noris2013} and Favreau et al. \shortcite{Favreau2016}. For their methods, we chose the best results out of all those options on a per-input basis. We run our method with default parameters on all inputs.

On the clean digital inputs, our results are comparable to the ones by Noris et al. (Fig.~\ref{fig:comparison_noris}, left). On the noisy inputs (Fig.~\ref{fig:comparison_noris}), right, our variational method reliably disambiguates junctions, even with missing details and varying stroke width. The method by Noris et al.~\shortcite{Noris2013} fails to disambiguate the complicated regions due to its discrete nature and heavy reliance on image gradients, e.g. puppy's eyes (red). Our method faithfully captures the principal directions and junctions even in those regions. We provide additional comparison results in the auxiliary materials.

We see the method by Favreau et al. \shortcite{Favreau2016} as complementary to our method: their method works best when significant simplification of the curve network is needed, while our goal is to stay true to the drawing even in the presence of noise (Fig.~\ref{fig:comparison_favreau}). For sketches with multiple overlapping strokes, our method aims to reconstruct all the single pen strokes, and in some cases it may not be the desired behavior (Fig.~\ref{fig:limitations} (c)). However, for those cases, our method may serve as a better input vectorization for further topological simplification \cite{Favreau2016,Simo-Serra2016}.

Our method is robust to changes in the input bitmap (Fig.~\ref{fig:robustness}), due in large part to its variational nature. Note that since drawings with multiple overlapping strokes, such as inputs in \cite{Favreau2016}, were not the focus of our work, we ran \cite{Bartolo2007} on the left image, while tracing the curves from the dark pixels in the original sketch.

\begin{figure}
	\includegraphics[width=\linewidth]{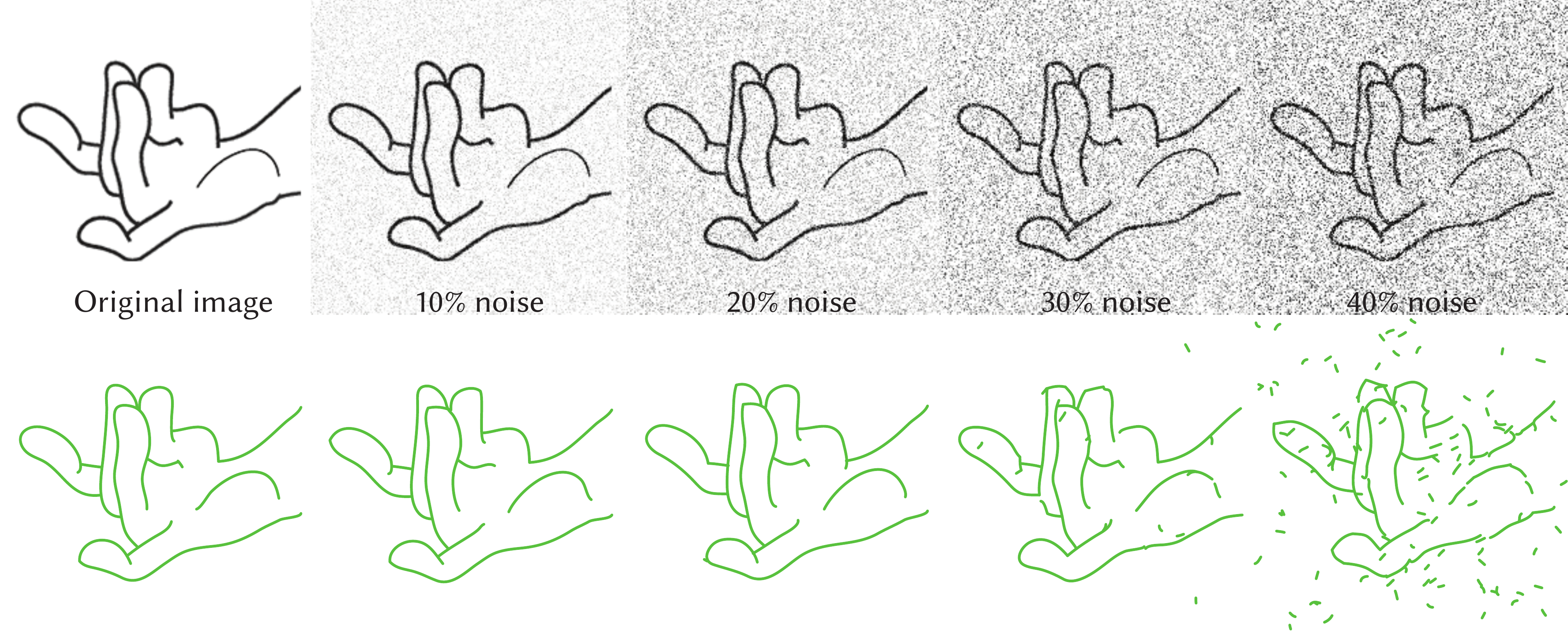}
	\caption{\revisedtext{Our method is robust under significant noise. Even with heavy noise, when discerning correct topology becomes problematic, junction accuracy remains stable.}}
	\label{fig:noise}
\end{figure}

\revisedtext{
	\paragraph{Noise robustness} We have evaluated noise robustness of our algorithm by testing on images polluted by various degrees of Gaussian noise (Fig.~\ref{fig:noise}). In general, our method is robust to Gaussian noise; junction directions are particularly stable.
	\paragraph{Benefits of individual steps} For completeness, we demonstrate effects of disabling significant steps of our algorithm in Figure \ref{fig:ablation_study}. 
}

\paragraph{Parameters and Processing Time} On a 4-core Intel i7-6700 @ 3.4Ghz with 32Gb RAM, our implementation usually takes from twenty seconds for lower-resolution images to a little over a minute on high resolution images. Due to the narrow-band optimization, our performance depends not on the image resolution, but rather on the number of dark pixels. Statistics for the images we tested are summarized in Table~\ref{table:time}. 
We use the same parameters for all the images: $\theta_{\mathrm{noise}}=0.35, n_\mathrm{hole}=4$.

While most parameters in our method have a straightforward and intuitive effect on the result, we have two main nonlinear weights: the frame field smoothness weight $\lambda$, and the regularizer weight $\mu$. In Figure~\ref{fig:changing_parameters}, we demonstrate that our method produces reasonable output under significant variations of $\lambda$. Namely, increasing the weight sharpens the junctions, at a possible cost of losing some fine details in the drawings. Therefore, higher values of $\lambda$ may be appropriate for noisier drawings. Our method also produces reasonable outputs for significantly different values of $\mu$ (Fig.~\ref{fig:changing_mu}).

\revisedtext{While we kept parameters fixed, $n_{hole}$ could be adjusted for drawings of different resolutions or noise structure. One could devise a heuristic to decrease it for lower resolutions, for instance, or use machine learning to calculate an optimal parameter for a class of drawings; we leave this for future work. We did not observe that keeping $n_{hole}$ fixed caused any issues in our experiments.}
\paragraph{Limitations} 
\begin{figure}
	\includegraphics[width=\linewidth]{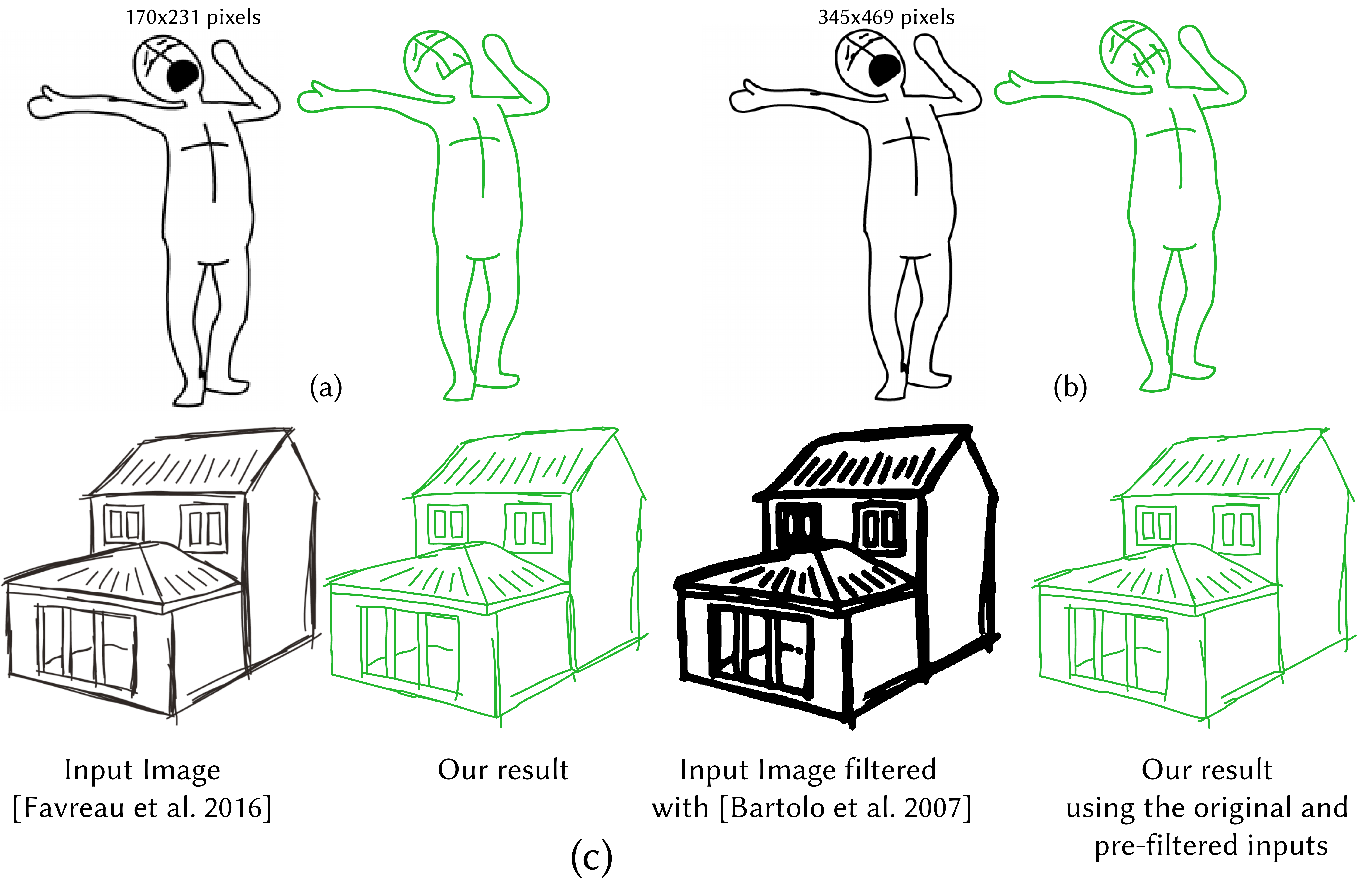}
	\caption{While our method is robust even for low-resolution images, similarly to most methods in the category, we do not support shaded areas (a,b). Input image \textcopyright Ksenia Popova. Since our method is aimed at truthfully vectorizing images, rough sketches with multiple overlapping strokes may require additional simplification (c). \revisedtext{To a certain degree, we can merge nearby parallel strokes by pre-filtering the input image with \cite{Bartolo2007} ((c), right), but tracing from the original narrow band, similarly to Fig.~\ref{fig:robustness}.}}
	\label{fig:limitations}
\end{figure}
Similarly to most methods of this category \cite{Noris2013,Favreau2016,Bo2016}, our method works best on drawings without shading (Fig.~\ref{fig:limitations}(a,b), the nose of the cat in Fig.~\ref{fig:robustness}). On shaded images, user interaction might be necessary to achieve correct vectorization. Very low resolution images might be also challenging to vectorize (Fig.~\ref{fig:changing_resolution}).

\begin{figure}
	\includegraphics[width=\linewidth]{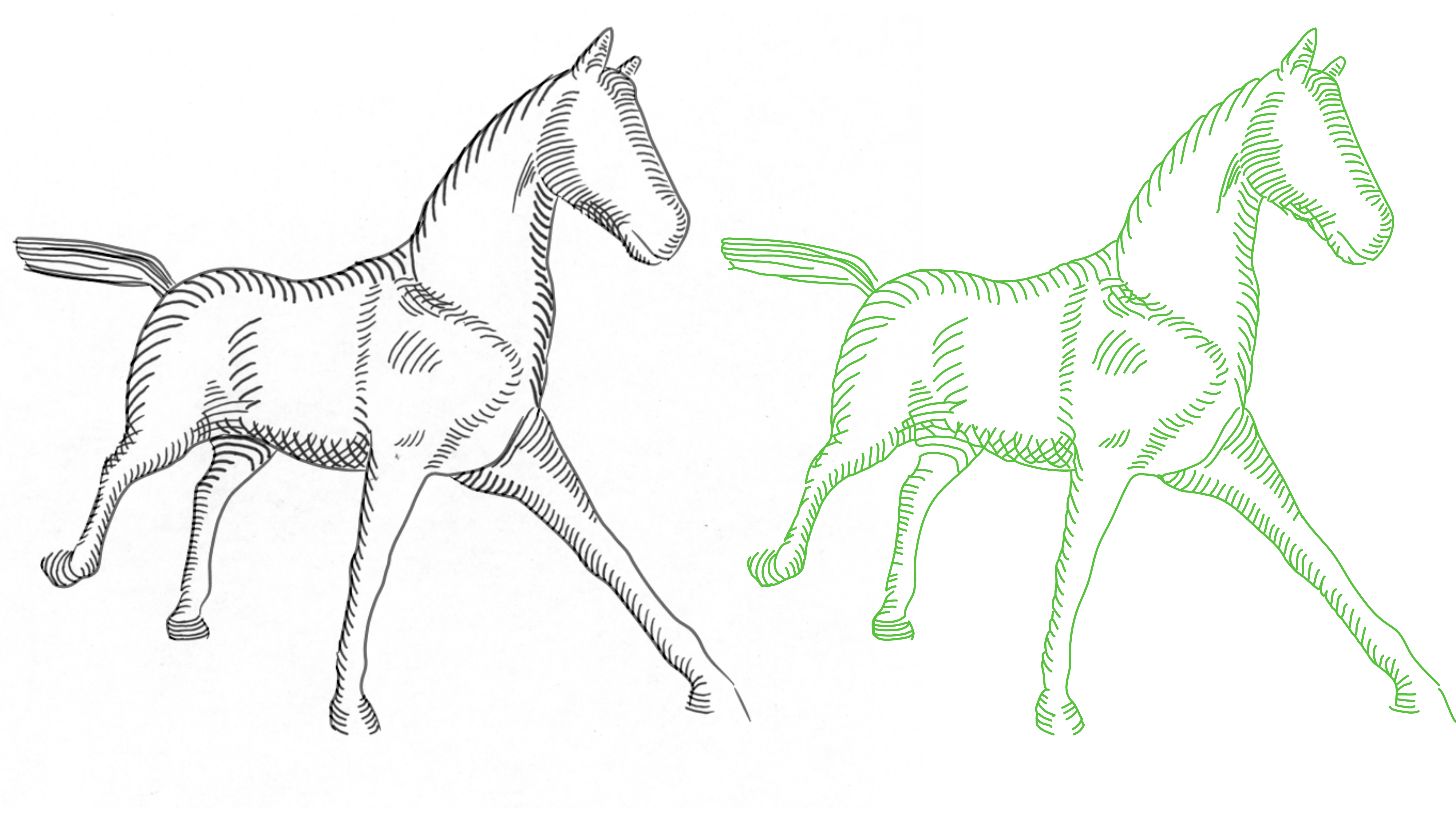}
	\caption{\revisedtext{Apart from line drawings, our method can be used to vectorize drawings in cross-hatching technique, when at most points there are only two curves crossing. Input drawing by Olga Vesselova \cite{Kalogerakis2012}.}}
	\label{fig:hatching}
\end{figure}

\revisedtext{A common alternative to shading is to convey information about shape and lighting via hatching \cite{Kalogerakis2012}. Those drawings are quite different from typical line drawings: in a technique called cross-hatching, artists would often draw ink strokes of three or more distinct hatching directions in one region. Since our frame field captures only two directions at a point, our method is not designed to vectorize natural hatching images. Even so, our method is naturally suited for vectorizing hatching examples when mostly only two sets of directions are used in every region (Fig.~\ref{fig:hatching}).} 

\section{Conclusion and Future Work}

We have presented a novel method for automatically vectorizing raster images, based on PolyVector field design. As we demonstrate on a gallery of examples, it reliably and efficiently disambiguates T- and X-junctions in both clean and noisy drawings, while staying true to curve shapes and connectivity. Our pipeline finds immediate application in artistic and engineering workflows, automatically providing a high-quality tracing without oversimplification or noise.

The presented method can be naturally extended to image domains where high-valence junctions are common, such as creating maps from GPS traces, by raising the degree of the PolyVector field polynomial (Eq. \ref{eq:complex_poly}). Our preliminary experiments indicate it is indeed a promising direction, but special care must be taken to find consistent matchings of the frame field roots in the presence of noise.

Another interesting potential extension is to vectorization of animated sequences: while a temporal coherence term is trivial to add in our frame field design framework, tracing might need to be modified to avoid temporal artifacts. %make the resulting sequence temporally coherent.

\section{Acknowledgments}
The authors acknowledge the generous support of Army Research Office grant W911NF-12-R-0011 (``Smooth Modeling of Flows on Graphs''), from the MIT Research Support Committee (``Structured Optimization for Geometric Problems''), and from the Skoltech--MIT Next Generation Program (``Simulation and Transfer Learning for Deep 3D Geometric Data Analysis'').

\bibliographystyle{ACM-Reference-Format}
\bibliography{polyvectorization}

\end{document}